\documentclass[aps,pre,preprint,superscriptaddress]{revtex4-1}  % for double-spaced preprint
\usepackage{graphicx}
\usepackage{subcaption}
\usepackage{dcolumn}
\usepackage{amssymb}
\usepackage{amsmath}
\usepackage{newtxtext,newtxmath,lipsum}
\usepackage[shortlabels]{enumitem}

\newcommand{\sech}{\normalfont\mbox{sech}\,}
\newcommand{\e}{\normalfont\mbox{e}\,}
\begin{document}
	\title{Baseband Modulational Instability and Manifestation of Breather and Super Rogue Wave Phenomena in Modified Noguchi Electrical Transmission Line}
	
	\author{N. Sinthuja}
	\affiliation{Department of Physics, Anna University, Chennai - 600 025, Tamilnadu, India}
	
	\author{M. Senthilvelan}
	\affiliation{Department of Nonlinear Dynamics, Bharathidasan University, Tiruchirappalli - 620 024, Tamilnadu, India}

    \author{K. Murali}
	\affiliation{Department of Physics, Anna University, Chennai - 600 025, Tamilnadu, India}
    
	\email{sinthukum14@gmail.com}
	
	\begin{abstract}
		\par  In this paper, we report the emergence of breather and super rogue waves in a modified Noguchi electrical transmission line and demonstrate that both phenomena arise from baseband modulational instability. We then systematically examine the influence of three key parameters and time that appear in the solutions, on the structure and behavior of breathers and super rogue waves, how they control waveform amplitude, localization, and intensity.  We also explore the intricate dynamics of wave propagation in nonlinear electrical line. Our findings provide valuable insights into the role of network parameters, such as inductances, capacitances, and nonlinear coefficients, in shaping wave behavior, offering not only practical guidelines for the design, stability, and experimental investigation of electrical transmission systems but also provides practical approaches for controlling and managing breather and super rogue wave phenomena in the transmission lines.
	\end{abstract}
	
	\maketitle

\section{Introduction}
Solitons, breathers, and rogue waves (RWs) are, key solutions in nonlinear wave dynamics, observed in fields like optics, oceanography, and Bose-Einstein condensates \cite{11,2,1}. Solitons are stable, localized waves that maintain their shape due to a balance between dispersion and nonlinearity. Breathers are oscillatory structures that change in amplitude and width over time. Rogue waves are rare, large-amplitude waves that emerge suddenly and vanish quickly \cite{Ankiewicz2016infinite,Wang2016breather,pelinov}. These waves result from modulational instability (MI) of wave packets and are studied in detail in the topics, say oceanography, optics, and plasma physics \cite{2,1,pelinov, Solli2007optical,Onorato2006extreme,Pelinovsky2004physical,Wang2018breather,Chabchoub2012super,21,22,Kengne2024coupled}. The MI has been proven to be necessary but not sufficient condition for RW formation. The sufficient condition for RW generation being the baseband MI. We define the baseband MI as the condition under which a continuous wave background is unstable with respect to perturbations having infinitesimally small frequencies \cite{Baronio2015baseband}. Baseband MI is a subset of MI that occurs in the low-frequency region, that is near to the baseband frequencies. It is important to study this instability because in this region modulations in the waves can grow leading to phenomena like RWs \cite{Baronio2015baseband,Narhi2016real,Yin2019optical,Baranio2014vector,Liu2023formation}. Extensive theoretical, numerical, and experimental studies were conducted on the aforementioned localized wave solutions to understand their behavior\cite{11,2,1,Ankiewicz2016infinite,Wang2016breather,Wang2018breather,Chabchoub2012super,21,22,Baronio2015baseband,Narhi2016real,Yin2019optical,Baranio2014vector,Liu2023formation}. It has been demonstrated that nonlinear Schr\"odinger (NLS) equation admits these three localized waves as solutions \cite{11,2,1}.

\begin{figure*}[!ht]
	\begin{center}
		\includegraphics[width=\linewidth]{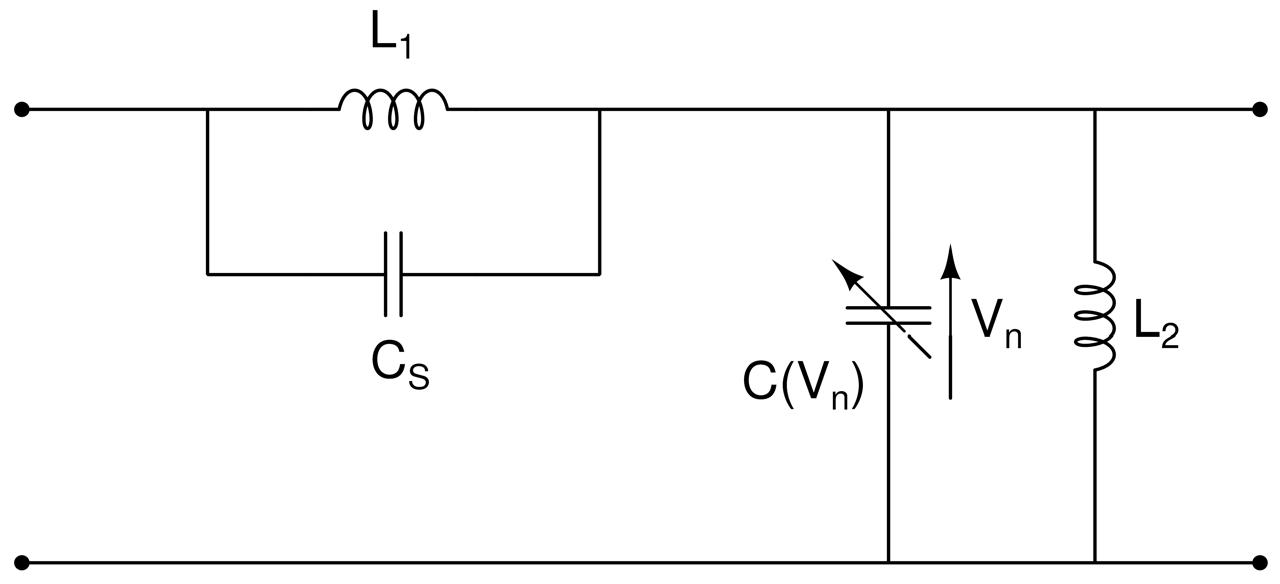}
		%\caption{}
	\end{center}
	\caption{Schematic diagram of a discrete nonlinear electrical transmission line cell}
	\label{fig01}
\end{figure*}

Nonlinear transmission lines serve as effective tools for studying wave propagation in nonlinear, dispersive media \cite{Kengnebook}. The electrical circuit shown in Fig. \ref{fig01}, which includes a linear dispersive capacitance $C_S$, differs from the Hirota and Suzuki model \cite{Hirota1970studies}. This distinctive network was developed by Noguchi specifically to study the propagation of first-order Korteweg-de Vries solitons in an experimental setting \cite{Noguchi1974solitons}. Unlike standard models, the inclusion of $C_S$ enables more accurate simulation of wave dispersion effects within electrical transmission lines, see for example Refs. \cite{Ichikawa1976contribution,Yoshinaga1984second,Marlund2006shukla,Kengne2006exact}.  However, most studies involving Noguchi's transmission line have not accounted for the effects of $C_S$. Pelap et al. have considered a modified Noguchi model to study the effect of linear capacitance $C_S$ on wave characteristics, addressing the limitations of traditional transmission lines in capturing nonlinear phenomena like solitons and MI in high-frequency applications \cite{Pelap2015dynamics}. Subsequently, the modified Noguchi's model is widely utilized to explore the effects of capacitance and inductance on soliton formation \cite{Kengne2017modelling,Kengne2019transmission}. Dissipative elements, such as resistance and conductance have been incorporated in some variants of the modified Noguchi model in order to represent real-world energy loss in transmission media \cite{Kengne2015dynamics,Kengne2022ginzeburg}. The key difference in the outcome of Noguchi and modified Noguchi model is that, without $C_S$, wave velocity is influenced only by nonlinear effects, whereas with $C_S$, both nonlinear and linear capacitance modify the wave velocity. In the original model, the wave shape is determined only by nonlinearity, whereas in the modified model, $C_S$ introduces dispersive effects, which can alter the wave shape. Additionally, soliton stability is assumed in the original model but is affected by $C_S$ in the modified model, which accounts for high-frequency dispersive effects, providing a more accurate representation of wave propagation \cite{Noguchi1974solitons,Pelap2015dynamics}.

In the beginning, solitons have attracted considerable interest in the context of electrical transmission lines due to their potential applications in the transmission stability \cite{Kengne2017modelling,Kengne2019transmission, Marquie1994generation}. Solitons, with their inherent stability and self-reinforcing properties, have been analyzed in various transmission line models, where they can propagate without losing energy over long distances \cite{3,31}. This behavior makes solitons promising for reliable, low-loss transmission in communication networks. Djelah et al. constructed first- and second-order RW solutions within a nonlinear electrical transmission line, incorporating an external potential and next-nearest-neighbor coupling in the modified Noguchi line \cite{Djelah2023first}. Additionally, super RW solutions for coupled electrical transmission lines have been investigated in \cite{Duan2020super}. A super RW is a subtype of RW that is exceptionally large, often exceeding five times the significant wave height.  

Understanding the emission of RWs in nonlinear transmission lines has become essential for mitigating such risks and optimizing line designs \cite{Kengne2020engineering}. The concepts of breathers and baseband MI have been overlooked in the context of the modified Noguchi electrical transmission lines. Furthermore, the study of super RW concept has not been addressed within the modified Noguchi electrical transmission line framework. To address this gap, our work primarily focuses on constructing a theoretical framework that explores breather, super RW, and their associated baseband MI in the modified Noguchi model. Notably, the super RW solution which we consider in this work differs from previous works both in terms of the circuit equation \cite{Duan2020super} and solution \cite{Kengne2019transmission,Djelah2023first}, highlighting to fill the gap in the field. Super RWs and breathers are extreme waveforms that help to characterize nonlinear effects in circuits, particularly the interaction of high-frequency signals. These studies improve our understanding of energy propagation, dispersion, and nonlinearity in electrical systems, crucial for signal processing. Baseband MI can destabilize low-frequency signals, leading to distortion. Understanding this behavior enables one to mitigate these effects, ensuring reliable data transmission and enhancing the design, efficiency, and stability of modern electrical and communication systems.

To carry out the above tasks first we obtain a circuit equation. We then reduce the circuit equation to the NLS equation with a potential term using the reductive perturbation method \cite{Taniuti1969perturbation}. Next, we transform the latter equation to the standard NLS equation with the help of a transformation. We then focus on breather and super RW solutions of this NLS equation. From the known solutions of the NLS equation, we present solutions for the circuit equation. In addition to the above, we investigate how the system parameters and time affect both breather dynamics and the super RWs. Our study also brings out baseband MI in the context of electrical transmission lines which in turn provide new insights into the formation of breathers and super RWs. We present for the first time breather and super RW solutions of the modified Noguchi electrical transmission line model, emphasizing the crucial influence of network parameters on wave behavior. Our findings not only extend theoretical understanding but also offer practical guidance for managing nonlinear wave phenomena in electrical systems, paving the way for more stable and efficient transmission line designs.

The paper is organized as follows: In Section 2, we apply the reductive perturbation method in the semi-discrete limit to derive the standard NLS equation. In Section 3, we address the baseband MI in this system. In Section 4, we explore breather dynamics within the electrical transmission line. In Section 5, we analyze the super RW in the model. In Section 6, we examine the roles of additional network parameters in both breathers and super RWs. Finally, in Section 7, we provide a summary of our findings.

\section{Model and Circuit equation}
\subsection{Formulation of circuit model and derivation of governing equation}
The model under investigation is a one-dimensional electrical transmission line characterized by both dispersive and nonlinear properties. The circuit given in Fig. \ref{fig01}, represents just one cell. The electrical line consists of $N$ identical cells arranged in sequence to form a continuous line. In each cell, there is a linear inductor, denoted as $L_1$, connected in parallel with a linear capacitor $C_S$ as shown in Fig. \ref{fig01}. This pairing enables the cell to exhibit linear inductive and capacitive behavior, supporting the transmission and dispersion of electrical signals. Additionally, this parallel configuration of $L_1$ and $C_S$ is connected in series with a secondary parallel configuration comprising another linear inductor $L_2$ and a nonlinear capacitor $C$. The nonlinear capacitance $C$ is implemented using a reverse-biased diode, which provides a voltage-dependent capacitance, allowing the circuit to respond nonlinearly to input voltages. The capacitance $C(V_n + V_b)$, which varies with the voltage $V_n$ across the $n$-th capacitor, is assumed to follow a nonlinear relationship. This nonlinear dependence on voltage can be effectively modeled using a polynomial expansion, enabling an approximation that captures the capacitance's variation with voltage more precisely, as discussed in detail in \cite{Djelah2023rogue}
%The model under consideration is a one-dimensional, dispersive, and nonlinear electrical transmission line comprising $N$ identical cells. Each cell includes a linear inductance $L_1$ in parallel with a linear capacitance $C_S$. This arrangement is then connected in series with a parallel combination of a linear inductance $L_2$ and a nonlinear capacitance $C$ formed by a reverse-biased diode. The differential capacitance $C(V_n + V_b)$ is assumed to be a nonlinear function of the voltage $V_n$ across the $n$-th capacitor. This capacitance-voltage relationship can be approximated by a polynomial expansion, as detailed in \cite{Djelah2023rogue}.
\begin{equation}
C(V_b+V_n)=\frac{dQ_n}{dV_n}=C_0(1-2\alpha V_n+3\beta V_n^2).
\label{eq1}
\end{equation}
Here, $C_0$ represents the characteristic capacitance, corresponding to the value of the nonlinear capacitance at the biasing voltage $V_b$. The constants $\alpha$ and $\beta$ are positive coefficients that describe the nonlinear behavior of the system. $\alpha$ characterizes the first-order nonlinearity, while $\beta$ represents the second-order nonlinearity, both of which influence the charge $Q_n$ stored in the $n$-th capacitor.

By applying Kirchhoff's laws, we obtain the following set of discrete differential equation that describe the wave propagation along the transmission line, that is
\begin{align}
\frac{d^2V_n}{dt^2}+u_0^2(2V_n-V_{n-1}-V_{n+1})&-\alpha\frac{d^2V_n^2}{dt^2}+\omega_0^2V_n+\beta\frac{d^2V_n^3}{dt^2}\nonumber\\&+\gamma \frac{d^2(2V_n-V_{n-1}-V_{n+1})}{dt^2}=0,
\label{eq2}
\end{align}
with $u_0=(L_1C_0)^{-\frac{1}{2}}$, $\omega_0=(L_2C_0)^{-\frac{1}{2}}$ and the parameter $\gamma=\frac{C_S}{C_0}$ represents the dispersive effect in the system. 

To analyze the dynamics of wave propagation within the network, we apply the reductive perturbation method in the semi-discrete limit \cite{Taniuti1969perturbation}. In this approach, we focus on excitations whose envelope changes slowly between adjacent cells, allowing one to approximate the electrical line as a continuous medium rather than as a discrete system. This transition from discrete to continuous behavior simplifies the analysis of wave dynamics in the network.

To move forward, we introduce the slow variables $\zeta=\epsilon(n-v_g t)$ and $\tau=\epsilon^2 t$, where $v_g$ represents the group velocity, $\epsilon$ is a small parameter, and $n$ denotes the cell number. These variables capture the slow evolution of the wave's envelope as it propagates through the network. By adopting this scaling, we effectively smooth out rapid variations between cells, making the system easier to model and analyze as a continuous medium.

\subsection{From the circuit model to the NLS equation with potential term}
Let us assume that the solution of Eq. (\ref{eq2}) in the following form \cite{Kengne2020engineering}:
\begin{equation}
V_n(t)=\epsilon\psi(\zeta,\tau)e^{i\theta}+\epsilon^2\psi_{10}(\zeta,\tau)+\epsilon^2 \psi_{20}(\zeta,\tau)e^{2i\theta}+c.c.
\label{eq3}
\end{equation}
Here, $\theta=(kn-\omega t)$ represents the phase that varies rapidly with both position and time, where $k$ is the wavenumber and $\omega$ is the angular frequency. The term 'c.c.' refers to the complex conjugate of the preceding expression. To incorporate the imbalance in the charge-voltage relationship described by Eq. (\ref{eq1}), both the DC component  $\psi_{10}(\zeta,\tau)$ and the second harmonic term $\psi_{20}(\zeta,\tau)$  are included along with the fundamental term $\psi(\zeta,\tau)$.

To begin, we substitute the assumed form of the solution, given in Eq. (\ref{eq3}), into the original circuit equation (\ref{eq2}). After this substitution, we focus on isolating and keeping terms that are proportional to different orders of $\epsilon$, as well as the exponential terms like $e^{i\theta}$, where $\theta$ is the rapidly varying phase. 

For considering the terms involving $(\epsilon, e^{i\theta})$ and solving the resultant equation, we obtain the following linear dispersion relation
\begin{equation}
\omega=\sqrt{\frac{\omega_0^2+4u_0^2\sin^2(\frac{k}{2})}{1+4\gamma\sin^2(\frac{k}{2})}}.
\label{eq5}
\end{equation}
From this we find the expression for group velocity in the form
\begin{equation}
v_g=\frac{d\omega}{dk}=\frac{(u_0^2-\gamma\omega^2)\sin(k)}{\omega(1+4u_0^2\sin^2(\frac{k}{2}))}.
\label{eq6}
\end{equation}
Similarly, for $(\epsilon^4, e^{i0})$, the following partial differential equation can be obtained
\begin{equation}
\frac{\partial^2 \psi_{20}}{\partial \zeta^2}=\frac{2\alpha v_g^2}{v_g^2-u_0^2}\frac{\partial^2|\psi|^2}{\partial \zeta^2}.
\label{eq7}
\end{equation}
\begin{figure*}[!ht]
	\begin{center}
		\includegraphics[width=\linewidth]{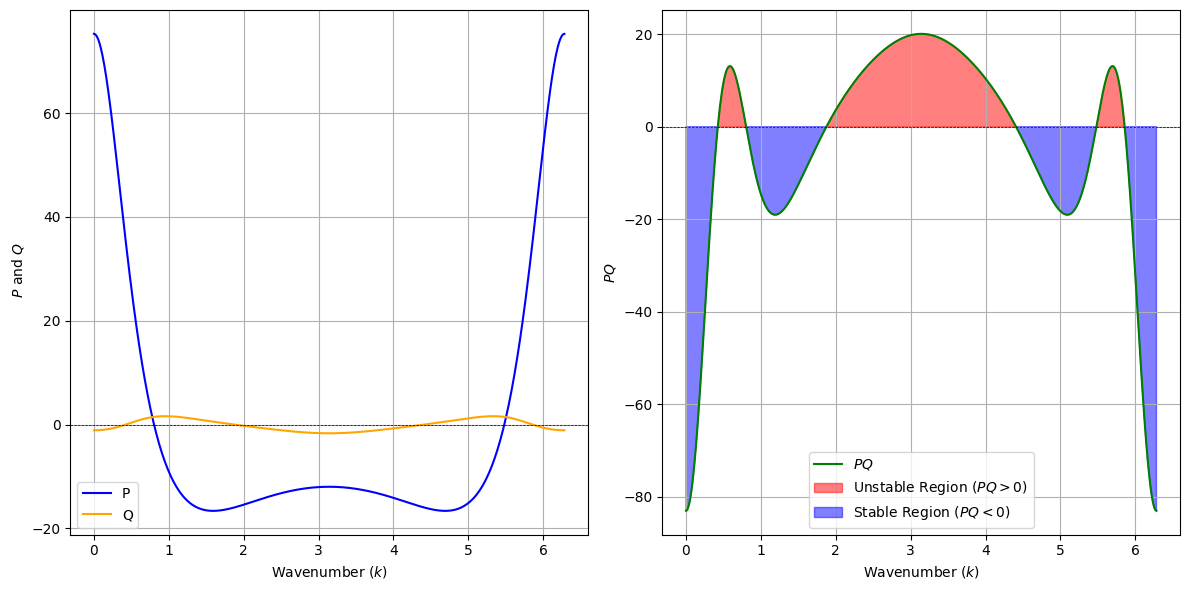}
		%\caption{}
	\end{center}
	\vspace{0.1cm}
	\caption{Dispersion and nonlinearity behaviour based on a wavenumber: (a) P and Q vs k, (b) PQ vs k.}
	\label{fig1}
\end{figure*} 
Solving Eq. (\ref{eq7}), we find
\begin{equation}
\psi_{10}(\zeta,\tau)=\frac{2\alpha v_g^2|\psi|^2}{v_g^2-u_0^2}+c_0(\tau)\zeta+c_1(\tau).
\label{eq8}
\end{equation}
Here the $c_0$ and $c_1$ are arbitrary functions of $\tau$. 

When considering terms involving $(\epsilon^2, e^{2i\theta})$, we can obtain an expression for the second harmonic term in the form
\begin{equation}
\psi_{20}(\zeta,\tau)=\frac{4\alpha\omega^2\psi^2}{4\omega^2+ 4(4\gamma\omega^2-u_0^2)\sin^2(k)-\omega_0^2}
\label{eq81}.
\end{equation}

Finally, from the terms involving $(\epsilon^3, e^{i\theta})$, we can get the following differential equation \cite{Kengne2015dynamics}
\begin{subequations}
	\label{eq9}
	\begin{equation}
	i\frac{\partial \psi}{\partial \tau}+P\frac{\partial^2\psi}{\partial \zeta^2}+Q|\psi|^2\psi+\Gamma(\tau)\zeta\psi=0,
	\label{eq9a}
	\end{equation} 
	in which
	\begin{align}
	P&=\frac{1}{1+4\gamma\sin^2(\frac{k}{2})}\Bigg(-\frac{v_g^2}{2\omega}\left(1+4\gamma \sin^2(\frac{k}{2})\right)+\left(\frac{u_0^2}{2\omega}-\frac{\gamma \omega}{2}\right)\cos(k)-2\gamma v_g \sin(k)\Bigg),\\
	Q&=\frac{\omega}{1+4\gamma\sin^2(\frac{k}{2})}\Bigg(\frac{3\beta}{2}-\frac{2\alpha^2v_g^2}{v_g^2-u_0^2}-\frac{4\alpha^2\omega^2}{4\omega^2-\omega_0^2+4(4\gamma\omega^2-u_0^2)\sin^2(k)}\Bigg),\\
	\Gamma(\tau)&=-\frac{\alpha\omega}{1+4\gamma\sin^2(\frac{k}{2})}c_0(\tau).
	\label{eq10}
	\end{align}
\end{subequations}
Equation (\ref{eq9a}) reduces to the NLS equation when $\Gamma(\tau)=0$. In our analysis we focus on the case where $\Gamma(\tau)$ is a non-trivial function. Equation (\ref{eq9a}), under the condition $\Gamma(\tau)\neq0$, is referred to as the Gross-Pitaevskii equation with a linear potential \cite{Agosta2000stationary}. Here, the dispersion coefficient $P$ describes the group velocity dispersion, whereas the self-modulation coefficient $Q$ measures the strength of the standard nonlinearity. 
\subsection{Transformation to the standard NLS equation}
The NLS equation is integrable when we apply scaling to it, and it admits several kinds of solutions including soliton, breather and RWs. To transform Eq. (\ref{eq9}) into NLS equation, we consider the following transformation,
\begin{subequations}
	\label{eq11}
	\begin{equation}
	\psi(\zeta,\tau)=\rho \Psi(X,T)e^{i(\lambda_1 \zeta+\lambda_2)},
	\end{equation}
	with
	\begin{align}
	X=\rho\zeta \sqrt{\frac{Q}{2P}}-V T,\quad T=\frac{1}{2}Q\rho^2\tau,\quad \frac{d\lambda_1}{d\tau}=\Gamma(\tau),\nonumber\\ \frac{d\lambda_2}{d\tau}=-P\lambda_1^2, \quad V=\frac{4\lambda_1P}{\rho Q}\sqrt{\frac{Q}{2P}},
	\label{ex1}
	\end{align}
\end{subequations}
where $\rho$ is a real constant. Substituting the above expressions (\ref{eq11}) into Eq. (\ref{eq9}) and rearranging, we end up at
\begin{equation}
i\frac{\partial \Psi}{\partial T}+\frac{\partial^2 \Psi}{\partial X^2}+2|\Psi|^2\Psi=0,
\label{eq12}
\end{equation}
which is nothing but the standard NLS equation. To determine $\lambda_1$ and $\lambda_2$ one has to solve the third and fourth equations of (\ref{ex1}). To proceed further we consider $c_0(\tau)$ in $\Gamma(\tau)$ as a real constant, that is $c_0(\tau)=c_0$. Solving Eq. (\ref{ex1}) with (\ref{eq10}), where $c_0(\tau)=c_0$, $\lambda_1$ and $\lambda_2$ take the following forms
\begin{align}
\lambda_1&=-\frac{\alpha\omega}{1+4\gamma\sin^2(k/2)}c_0\tau+d_1,\\
\lambda_2&=-P\Bigg(\frac{\alpha^2\omega^2c_0^2}{3(1+4\gamma\sin^2(k/2))^2}\tau^3-\frac{\alpha\omega c_0 d_1}{1+4\gamma\sin^2(k/2)}\tau^2+d_1^2\tau\Bigg)+d_2,
\end{align}
where $d_1$ and $d_2$ are the integration constants.
\section{Baseband Modulational Instability}
\subsection{Definition of baseband MI in the electrical transmission line}
Baseband MI refers to a specific case where the instability occurs for perturbations at low frequencies, often near zero. In this case, the MI focuses on perturbations around the baseband (low-frequency range), which causes a self-modulation of the amplitude at low frequencies. In the context of a nonlinear transmission line, baseband MI can occur under the right conditions when the dispersion and nonlinearity lead to an imbalance. The baseband MI occurs when the perturbation frequencies grow exponentially in the low-frequency region. This type of instability is significant in applications such as signal processing in electrical transmission lines, where low-frequency perturbations can lead to large amplitude fluctuations \cite{Baranio2014vector,Baronio2015baseband}.
\begin{figure*}[!ht]
	\begin{center}
		\begin{subfigure}{0.45\textwidth}
			%\caption{}
			\includegraphics[width=\linewidth]{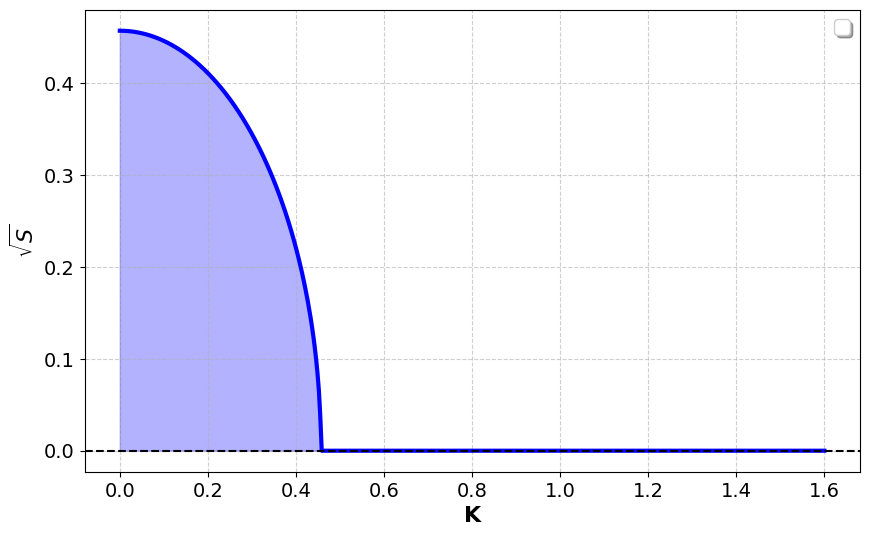}
			\caption{}
		\end{subfigure}
		\begin{subfigure}{0.45\textwidth}
			%\caption{}
			\includegraphics[width=\linewidth]{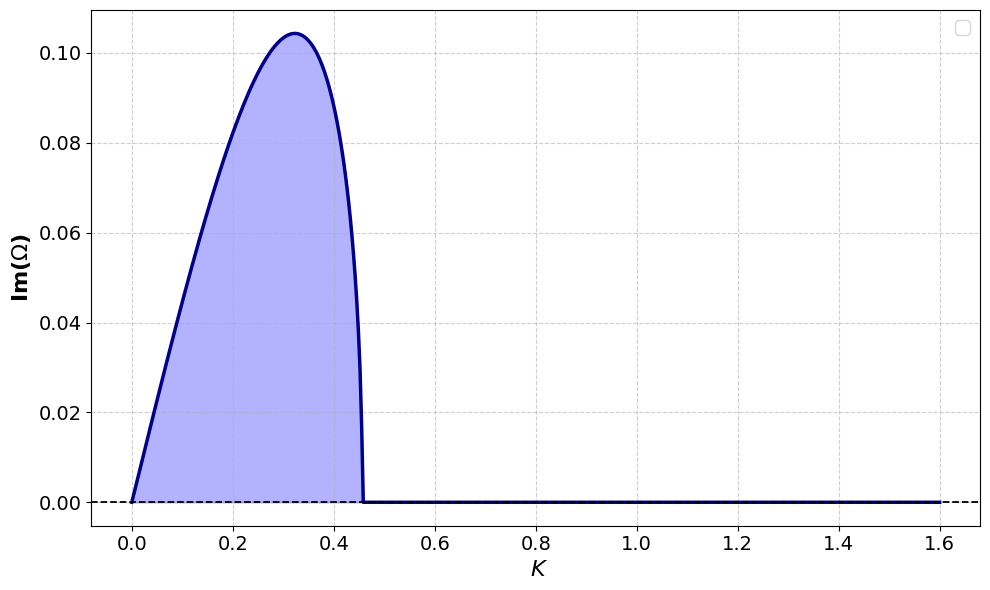}
			\caption{}
		\end{subfigure}
	\end{center}
	\vspace{-0.5cm}
	\caption{Instability growth rate, indicating the presence of Baseband MI. (a) $\sqrt{S}$ vs K, (b) $Im(\Omega)$ vs K}
	\label{fig2}
\end{figure*} 

To analyze baseband MI, we linearize the governing equation with a plane wave solution and solve the resulting dispersion relation. If the growth rate for perturbations at or near zero frequency is positive, it indicates that the system is prone to baseband MI. This phenomenon is crucial for understanding the behavior of signals in nonlinear media, including optical fibers and electrical transmission lines.  
\subsection{Perturbation analysis of plane wave solutions for the model (\ref{eq2})}
We begin by considering solutions of the governing equation (\ref{eq9}) for the nonlinear transmission line in the form of a plane wave with perturbations:
\begin{equation}
\psi(\zeta,\tau)=A_2 \;e^{i\left(\frac{\zeta^2+\mu\zeta}{4P(\tau+\tau_0)}-\int_{\nu_0}^{\nu}\tilde{\Omega}(y)dy\right)}.
\label{meq1}
\end{equation}
Here, $A_2=\frac{A_1}{\sqrt{a_0(\tau+\tau_0)}}$, $\frac{d\mu}{d\tau}=4P(\tau+\tau_0)\Gamma(\tau)$, $\nu=-\frac{P}{a_0^2(\tau+\tau_0)}$ and $\tilde{\Omega}(\nu)=\frac{a_0^2\mu^2}{16P^2}+\frac{QA_1^2}{a_0\nu}$. The above equation contains two arbitrary real constants, $\tau_0$ and $A_1$, and another constant, $a_0$ which is a real constant derived from the condition $a_0(\tau+\tau_0)>0$. The function $\tilde{\Omega}(\nu)$ represents the time-dependent nonlinear frequency shift. To investigate the MI condition of the carrier wave, we introduce a small disturbance $\delta\psi(\zeta,\tau)$ to the solution (\ref{meq1}) in the form
\begin{equation}
\psi(\zeta,\tau)=A_2(1+\delta\psi(\zeta,\tau)) \; e^{i\left(\frac{\zeta^2+\mu\zeta}{4P(\tau+\tau_0)}-\int_{\nu_0}^{\nu}\tilde{\Omega}(y)dy\right)}.
\label{meq2}
\end{equation}
Substituting Eq. (\ref{meq2}) into (\ref{eq9}) and retaining only the linear terms in $\delta\psi$ and its conjugate $\delta\bar{\psi}$, we obtain 
\begin{equation}
i\frac{\partial}{\partial \tau}\delta\psi+P\frac{\partial^2}{\partial \zeta^2}\delta\psi-i\frac{2\zeta+\mu}{2(\tau+\tau_0)}\frac{\partial}{\partial\zeta}\delta\psi+\frac{QA_1^2}{a_0(\tau+\tau_0)}(\delta\psi+\delta\bar{\psi})=0.
\label{meq3}
\end{equation}
This equation governs the evolution of the small perturbation $\delta\psi$, describing how it evolves in both time $\tau$ and space $\zeta$ under the influence of the nonlinearities. The first term represents the temporal evolution of the perturbation, the second term corresponds to the spatial dispersion, and the third term is a convective term that accounts for the interaction between the perturbation and the background. The final term models the nonlinear interaction between the perturbation and its complex conjugate, which is essential for the MI. To proceed, we consider the perturbation in a specific form that allows us to study its stability and determine the conditions for the onset of instability, which is given by:
\begin{equation}
\delta\psi=B_1\;\e^{i\left(\frac{K\zeta}{\tau+\tau_0}-\int_{\nu_0}^{\nu}\Omega(y)dy\right)}+\bar{B}_2\;\e^{i\left(-\frac{K\zeta}{\tau+\tau_0}+\int_{\nu_0}^{\nu}\bar{\Omega}(y)dy\right)},
\label{meq4}
\end{equation}
where $B_1$ and $B_2$ are complex constants. The parameter $K$ represents the wavenumber, and $\Omega$  denotes the pulsation (angular frequency) of the perturbation. Substituting Eq. (\ref{meq4}) into Eq. (\ref{meq3}) leads to the perturbation wavenumber-frequency relation:
\begin{equation}
\left[\Omega-\frac{K\mu a_0^2}{2P}\right]^2=K^2 a_0^4\left[K^2-\frac{2QA_1^2}{Pa_0}(\tau+\tau_0)\right].
\label{meq5}
\end{equation}
\begin{figure*}[!ht]
	\begin{center}
		\begin{subfigure}{0.45\textwidth}
			%\caption{}
			\includegraphics[width=\linewidth]{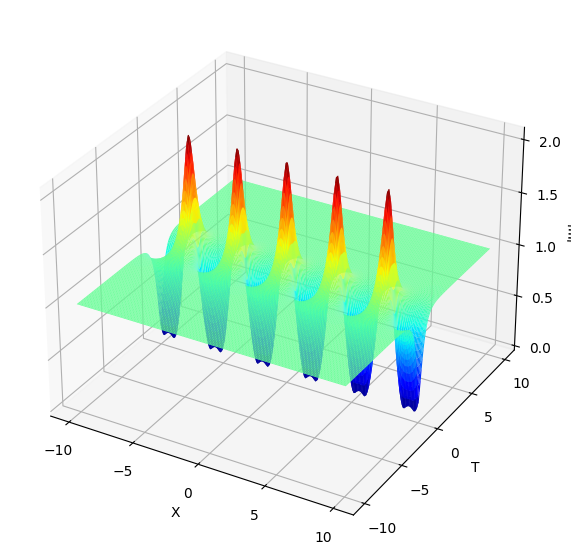}
			\caption{}
		\end{subfigure}
		\begin{subfigure}{0.45\textwidth}
			%\caption{}
			\includegraphics[width=\linewidth]{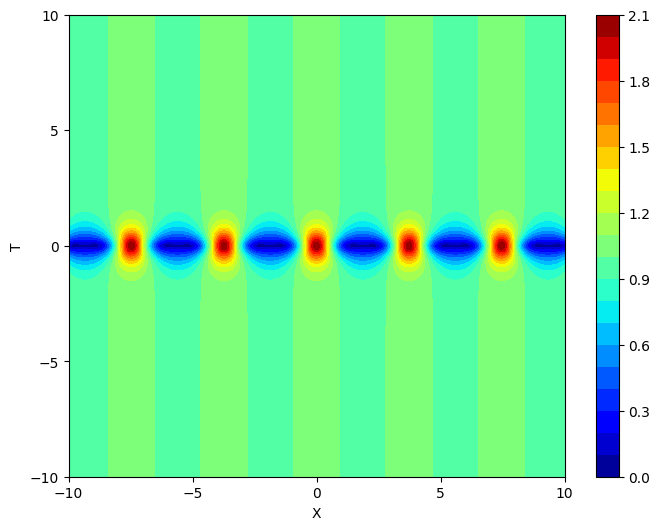}
			\caption{}
		\end{subfigure}
	\end{center}
	\vspace{0.1cm}
	\caption{The Akhmediev breather ($|\Psi_b (X,T)|$) for the NLS equation (\ref{eq12}) with $c=1$. (a) 3D plot and (b) contour plot.}
	\label{figb1}
\end{figure*} 
Here we assume that $a_0(\tau+\tau_0)>0$, and we immediately observe that the behavior of $\Omega$ for a given wavenumber $K$ is determined by the sign of $P/Q$, which influences whether the frequency shift increases or decreases, thus affecting the stability of the perturbation.
\subsection{Modulationally stable or unstable regions}
When $PQ<0$, the term inside the square root on the right-hand side of equation (\ref{meq5}) remains positive for all values of the wavenumber $K$. This ensures that the right-hand side remains real and non-negative, which implies that $\Omega$ is always a real function. A real $\Omega$ indicates that the perturbation does not grow exponentially, and the amplitude remains constant. Therefore, the plane wave is modulationally stable.

When $PQ>0$, the term inside the square root can become negative for certain values of $K$. In such a case, the right-hand side of equation (\ref{meq5}) becomes complex and in this case, we have
\begin{equation}
\Omega=\frac{a_0^2K\mu}{2P}\pm iK a_0^2\sqrt{\frac{2QA_1^2(\tau+\tau_0)}{Pa_0}-K^2}.
\label{meq6}
\end{equation}
The presence of a non-zero imaginary part in $\Omega$  leads to exponential growth of the perturbation, which corresponds to MI. The perturbation's amplitude grows in time, causing the plane wave to become unstable. The growth rate of MI is defined by
\begin{equation}
|Im(\Omega)|=a_0^2K\sqrt{\frac{2QA_1^2(\tau+\tau_0)}{Pa_0}-K^2}\;\; \text{or} \;\; a_0^2K \sqrt{S}.
\label{meq7}
\end{equation}
where $S=\sqrt{\frac{2QA_1^2(\tau+\tau_0)}{Pa_0}-K^2}$.

Kengne et al. have also shown that the system behaves as stable or unstable depending on the sign of the product $PQ$, and they discussed the underlying facts in \cite{Kengne2017modelling, Kengne2019transmission}. In our study, we specifically focus on baseband MI, where the same criterion of $PQ$ governs the transition between stable and unstable regimes. However, in the context of baseband MI, this instability is characterized by the growth of low-frequency perturbations, which is more sensitive to the value of $PQ$ in the baseband region compared to the broader system stability. Thus, here, we primarily focus on baseband MI with the graphical descriptions. 

Figure \ref{fig1} demonstrates the behavior of the functions $P$ and $Q$ (the parameters related to the stability analysis) and the product $PQ$, which helps to determine the modulational stability or instability of the system. Here, we consider the parameter values $L_1=220 mH$, $L_2=470 mH$, $C_0=370\mu F$,  $C_S=56\mu F$,  $\alpha=0.21V^{-1}$, $\beta=0.0197V^{-2}$ and $k$ ranging from $0$ to $2\pi$. In Fig. \ref{fig1}(a), the blue curve represents the function $P$, which is related to the group velocity dispersion and affects how energy is distributed across different frequencies (wavenumbers). This curve provides insight into how the wave propagation characteristics change with $k$. The orange curve represents the function $Q$, which is tied to the nonlinearity responsible for self-phase modulation. This term indicates the strength of the nonlinear effects as a function of $k$. The behavior of both $P$ and $Q$ indicates how the dispersive and nonlinear properties vary as a function of the wavenumber. When both functions are positive or negative simultaneously, they reinforce each other, leading to instability. In Fig. \ref{fig1}(b), the green curve shows the product $PQ$. The regions where $PQ$ is positive indicate MI, while the regions where $PQ$ is negative correspond to stability. The red-shaded regions represent the wavenumber ranges where $PQ>0$. These are the unstable regions where MI occurs. In these regions, the imaginary part of the frequency causes perturbations to grow exponentially, leading to instability in the plane wave. The blue-shaded regions represent the stable regions where $PQ<0$. Here, perturbations do not grow, and the plane wave remains stable.

The connection between baseband MI and the formation of breathers or RWs explains how seemingly stable conditions can lead to extreme wave events when small perturbations are amplified. Baseband MI refers to the case where long-wavelength perturbations (low-frequency fluctuations) become unstable, leading to the amplification of small disturbances near the wavenumber $K \approx 0$ (see Refs. \cite{Liu2023formation,Benjamin1967the}). This behavior is crucial in nonlinear wave systems, as it facilitates the emergence of localized rogue waves through the continuous growth of initial perturbations. The key characteristic that distinguishes baseband MI from passband MI is the onset of instability at $K \approx 0$ (wavenumber values close to zero, not exactly zero), whereas passband MI arises only at finite $K \gg 0$, giving rise to periodic structures known as breathers. Figure \ref{fig2} illustrates the MI growth rate as a function of the perturbation wavenumber $K$. The following parameter values were considered: $L_1 = 220mH$, $L_2 = 470mH$, $C_0 = 370\mu F$, $C_S = 56,\mu F$, $\alpha = 0.21V^{-1}$, $\beta = 0.0197 V^{-2}$, $k = 0.75$, $\tau = 1$, $\tau_0 = 0.2$, $A_1 = 1$, and $a_0 = 1$. In Fig. \ref{fig2}(a), the plot shows $\sqrt{S(K)}$ versus $K$, where $S$ is the expression inside the square root of the MI growth rate formula. Although $\Im(\Omega) = 0$ exactly at $K = 0$ from Eq. (\ref{meq7}), we separately plotted $\sqrt{S(K)}$, which represents the instability potential. This term reaches its maximum at $K = 0$, indicating that low-frequency perturbations (near-zero $K$) have the highest potential to grow under MI.  However, since the full growth rate $\Im(\Omega)$ includes a multiplicative factor of $K$, its maximum does not occur at $K = 0$ but at a small positive $K > 0$. This is typical for baseband MI: the instability band starts at $K = 0$, and $\Im(\Omega) > 0$ for a continuous range of small wavenumbers. This behavior is clearly shown in Fig. \ref{fig2}(b), confirming that the system exhibits baseband MI. The instability is primarily confined to low wavenumbers, which means the system amplifies long-wavelength, low-frequency modulations. Our results confirm that the system exhibits baseband-dominated MI, with the origin of instability beginning in the baseband. Therefore, we establish that both rogue waves and breathers in this system are formed based on baseband MI. One can also explore passband MI by selecting different parameter values, under which the maximum MI growth rate occurs at $K \gg 0$.

\section{Breather in an electrical transmission line}
\subsection{Breather solution for the NLS equations (\ref{eq12}) and (\ref{eq9})}
Baseband MI causes the initial growth of low-frequency perturbations on a plane wave background. This amplification can lead to breather solutions, which are localized, structured peaks that oscillate periodically. In certain cases, these breather formations may evolve further under MI, eventually collapsing into highly localized, high-amplitude RWs. Thus, breathers can be considered precursors to RW events. In the literature, MI, especially at the baseband, is recognized as a mechanism that not only leads to RWs but also gives rise to breather structures. Breathers that can emerge due to baseband MI include Akhmediev breathers, which oscillate spatially on a plane wave background and can peak in amplitude, potentially resembling RWs under intensified MI. In our work, we focus specifically on Akhmediev breathers to explore their unique properties and implications.

Akhmediev breather solution for the NLS Eq. (\ref{eq12}) reads \cite{Zhou2021deep}

\begin{equation}
\Psi_b(X,T)=\frac{\cosh(q_{1} T-2ic)-\cos(c)\cos(q_2X)}{\cosh(q_1 T)-\cos(c)\cos(q_2 X)}e^{2iT},
\label{eqb1}
\end{equation}
where $q_1=2\sin(2c)$, $q_2=2\sin(c)$ and $c$ is a real constant.

Using the above solution, we can derive the solution of Eq.(\ref{eq9}) through (\ref{eq11}), which in turn take the form
\begin{align}
\psi_b(\zeta,\tau)&=\rho\left(\frac{\cosh\left[\frac{q_1Q\rho^2}{2}\tau-2ic\right]-\cos[c]\cos\left[q_2\rho\sqrt{\frac{Q}{2P}}\right]}{\cosh\left[\frac{q_1Q\rho^2}{2}\tau\right]-\cos[c]\cos\left[q_2\rho\sqrt{\frac{Q}{2P}}\right]}\right) e^{i[(\lambda_1\zeta+\lambda_2)+Q\rho^2\tau]},
\label{eqb2}
\end{align}
where $q_1$, $q_2$, $Q$ and $P$ are already given in Eqs. (\ref{eqb1}) and (\ref{eq9}) respectively. The considered Akhmediev breather solution is different from the one given in Ref. \cite{Guy2018construction}. The model and the governing circuit equation considered in \cite{Guy2018construction} are also different. Therefore, our breather solution exhibits unique behavior.
\begin{figure*}[!ht]
	\begin{center}
		\begin{subfigure}{0.45\textwidth}
			%\caption{}
			\includegraphics[width=\linewidth]{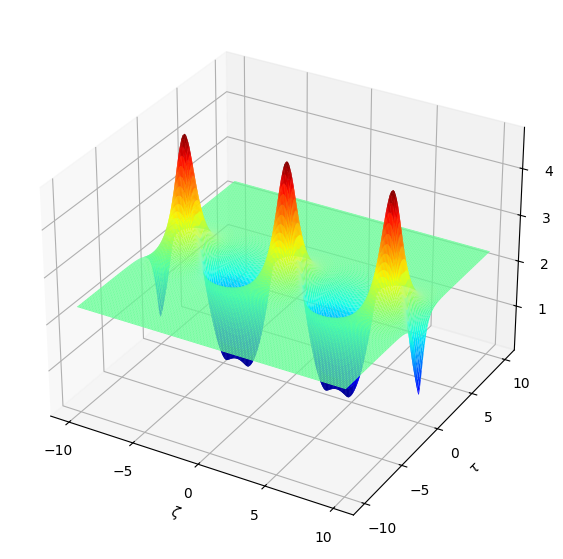}
			\caption{}
		\end{subfigure}
		\begin{subfigure}{0.45\textwidth}
			%\caption{}
			\includegraphics[width=\linewidth]{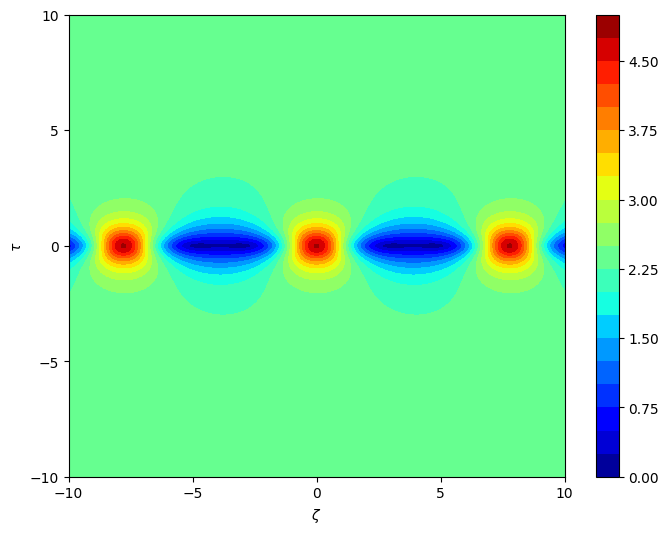}
			\caption{}
		\end{subfigure}\\\begin{subfigure}{0.45\textwidth}
			%\caption{}
			\includegraphics[width=\linewidth]{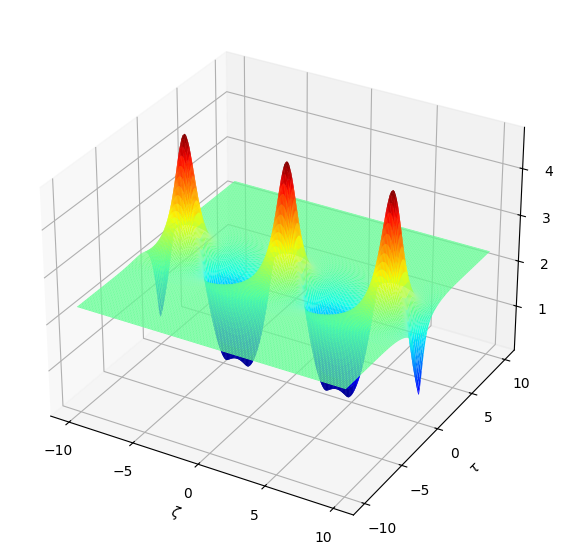}
			\caption{}
		\end{subfigure}
		\begin{subfigure}{0.45\textwidth}
			%\caption{}
			\includegraphics[width=\linewidth]{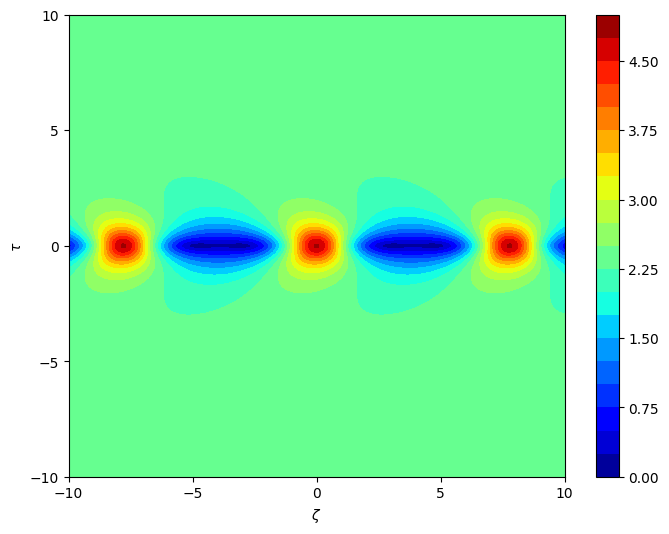}
			\caption{}
		\end{subfigure}
	\end{center}
	\vspace{0.1cm}
	\caption{The Akhmediev breather ($\psi_b(\zeta,\tau)$) for the NLS equation (\ref{eq9}) with $c=1$. (a), (b): 3D and contour plots for $c_0=0.0001$, (c), (d): 3D and contour plots for $c_0=0.001$.}
	\label{figb2}
\end{figure*}  

The Akhmediev breather for the Eq. (\ref{eq12}) is shown in Fig. \ref{figb1}. Figures \ref{figb1}(a) and \ref{figb1}(b) present the 3D plot and contour plot of the Akhmediev breather for the Eq. (\ref{eq12}), respectively. From these figures, we observe that the amplitude of the Akhmediev breather is approximately $|\psi(\zeta,\tau)| \approx 2.1$, with five peaks ($X$ range $-10$ to $10$) that oscillate periodically in space.

Figure \ref{figb2} presents the same information as Fig. \ref{figb1} but in experimental coordinates, indicating that the figure is plotted for the solution of Eq. (\ref{eq11}). For the numerical simulation of the plot, we utilize the parameters that accurately model the behavior of the system. The parameters are chosen based on standard or commonly used values in the relevant field, ensuring that the simulation reflects realistic conditions. By selecting these specific values for the wavenumber ($k=0.75$) inductance, capacitance, and other relevant variables, we can effectively study the dynamics of wave propagation or other phenomena under consideration and they are :
\begin{align}
L_1=220 mH,\; &L_2=470 mH, \; C_0=370\mu F, \; C_S=56\mu F, \; \alpha=0.21V^{-1}, \nonumber\\
\beta=0.0197V^{-2}.
\label{eq4}
\end{align}
The Akhmediev breather solution with experimental coordinates is presented as a 3D plot in Fig. \ref{figb2}(a), while the contour plot is shown in Fig. \ref{figb2}(b) for $c_0=0.0001$, $\rho=2.3$, $d_1=0$ and $d_2=0$. While the breather structure retains the same $\zeta$ range, notable differences can be observed in the experimental coordinate system: the peak amplitude has increased to approximately $|\psi(\zeta,\tau)| \approx 4.50$, and the number of observed peaks has decreased to three. Figures \ref{figb2}(c) and \ref{figb2}(d) correspond to Figs. \ref{figb2}(a) and \ref{figb2}(b), respectively, but for $c_0=0.001$. In Fig. \ref{figb2}(d), we can clearly see that the sky-blue region inside the contour plot has a loss compared to Fig. \ref{figb2}(b). These variations illustrate the impact of incorporating experimental coordinates on the characteristics of the Akhmediev breather. For higher $c_0$ values, the breathers exhibit a larger loss, meaning that their peak amplitude changed, their structure become less localized, or the number of observed peaks reduces.
\subsection{Breather solution in the modified Noguchi electrical transmission line}
To construct a breather solution of the considered electrical transmission line, we consider
\begin{equation}
V_{nb}(t)=\epsilon\psi_b(\zeta,\tau)e^{i\theta}+\epsilon^2\psi_{10}(\zeta,\tau)+\epsilon^2 \psi_{20}(\zeta,\tau)e^{2i\theta}+c.c.,
\label{eqb3}
\end{equation}
where $\psi(\zeta,\tau)$, $\psi_{10}(\zeta,\tau)$ and $\psi_{20}(\zeta,\tau)$ are given in Eqs. (\ref{eqb2}), (\ref{eq8}) and (\ref{eq81}). Equation (\ref{eqb3}) describes the dynamics of breathers in the electrical transmission line.

\begin{figure*}[!ht]
	\begin{center}
		\begin{subfigure}{0.45\textwidth}
			%\caption{}
			\includegraphics[width=\linewidth]{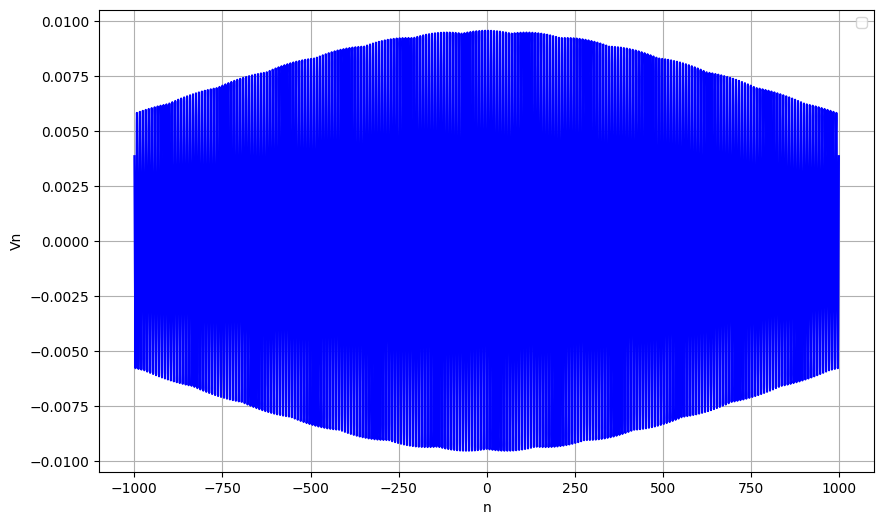}
			\caption{}
		\end{subfigure}
		\begin{subfigure}{0.45\textwidth}
			%\caption{}
			\includegraphics[width=\linewidth]{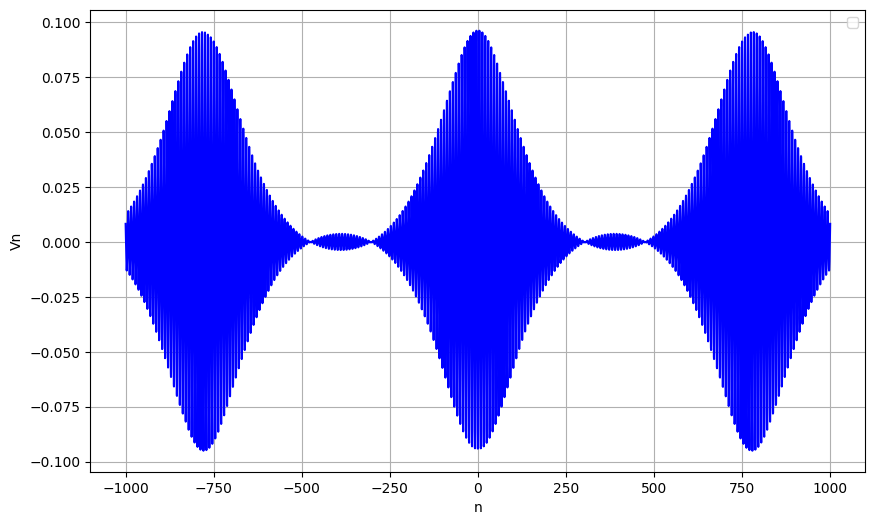}
			\caption{}
		\end{subfigure}\\
		\begin{subfigure}{0.45\textwidth}
			%\caption{}
			\includegraphics[width=\linewidth]{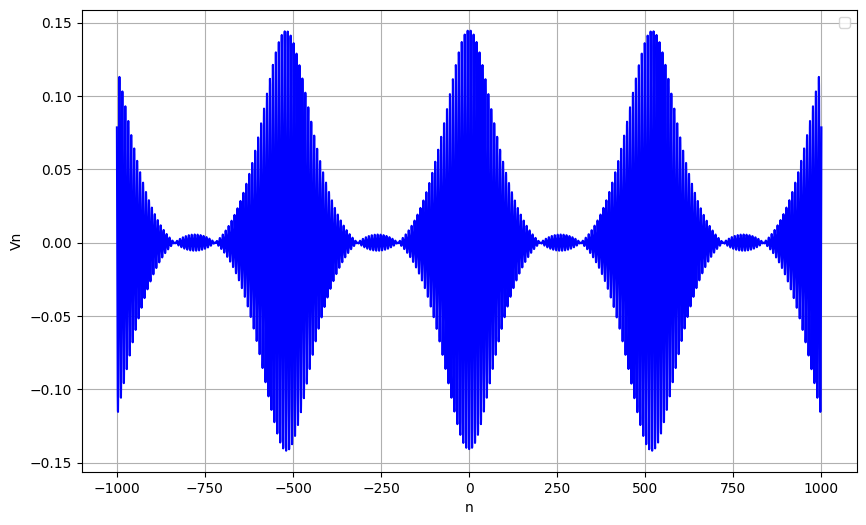}
			\caption{}
		\end{subfigure}
		\begin{subfigure}{0.45\textwidth}
			%\caption{}
			\includegraphics[width=\linewidth]{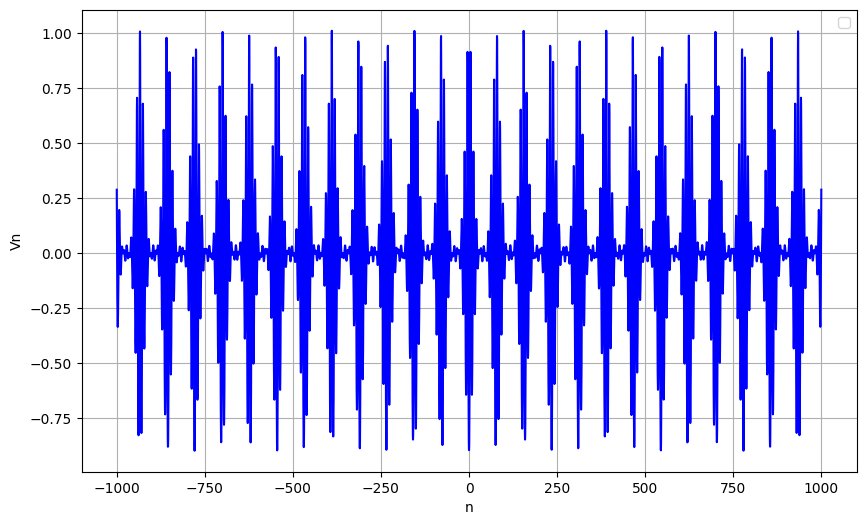}
			\caption{}
		\end{subfigure}
	\end{center}
	\vspace{-0.3cm}
	\caption{Variation of the Akhmediev breather in an electrical transmission line (\ref{eq2}) with parameter $\epsilon$ at $t=0$. (a) $\epsilon=0.001$; (b) $\epsilon=0.01$; (c) $\epsilon=0.015$; (d) $\epsilon=0.1$.}
	\label{figb3}
\end{figure*} 
Figure \ref{figb3} represents the Akmediev breather in the electrical transmission line (\ref{eqb3}) with varying values of parameter $\epsilon$ at $t=0$. The other parameters are $k=0.75$, $\rho=2.3$, $c_0 = 0.0001, d_1 = 0, d_2 = 0$ and $c_1$ is zero. As $\epsilon$ changes, the shape and amplitude of the breather adjust accordingly, demonstrating how variations in $\epsilon$ affect the breather profile while the remaining parameters are held constant at the specified values. In Fig. \ref{figb3}(a), for $\epsilon=0.01$, the initial formation of the breather wave is observed, though it is not fully developed in the region $n$ (cell number) from $-1000$ to $1000$. When  $\epsilon$ is increased to $0.01$, the full occurrence of the breather is clearly visible, as shown in Fig. \ref{figb3}(b). Further increasing the value of $\epsilon$ to $0.015$ and $0.1$ results in a change in the localization of breather peaks, as well as an increase in the amplitude and number of peaks (enhancement in the oscillatory behavior) in the breather, as shown in Fig. \ref{figb3}(c) and \ref{figb3}(d), respectively.
From these figures, we can see that as $\epsilon$ increases, the shape of the breather changes and the number of breather peaks and its amplitude increases. It also starts to oscillate more, creating more complex waveforms. These changes show how important $\epsilon$ is in affecting the breather's behavior, which could have a significant impact on signal quality and performance in transmission systems.

\begin{figure*}[!ht]
	\begin{center}
		\begin{subfigure}{0.45\textwidth}
			%\caption{}
			\includegraphics[width=\linewidth]{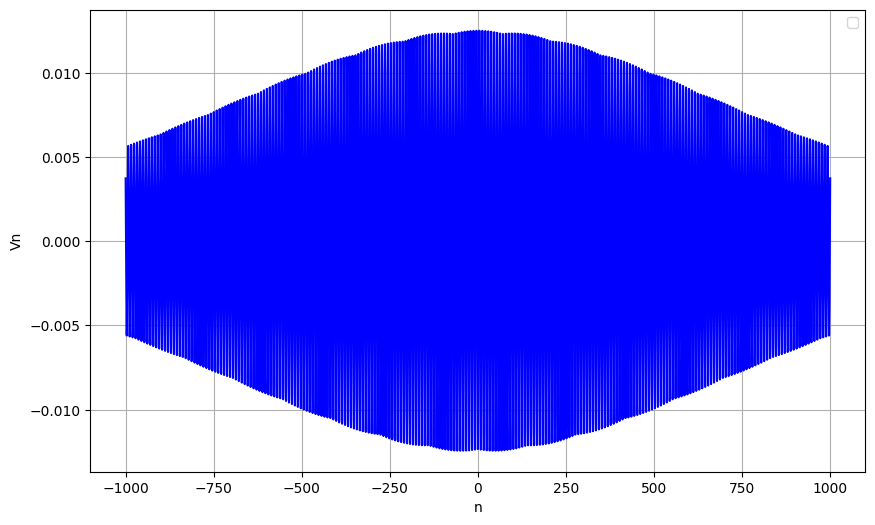}
			\caption{}
		\end{subfigure}
		\begin{subfigure}{0.45\textwidth}
			%\caption{}
			\includegraphics[width=\linewidth]{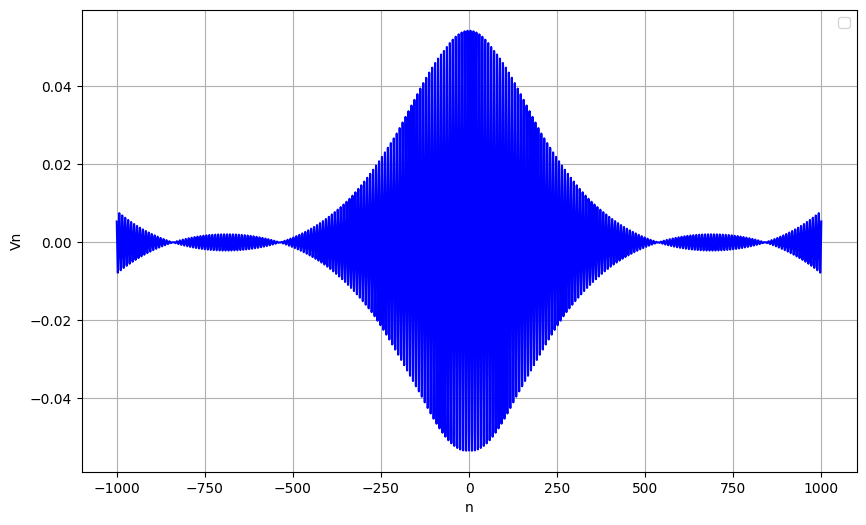}
			\caption{}
		\end{subfigure}\\
		\begin{subfigure}{0.45\textwidth}
			%\caption{}
			\includegraphics[width=\linewidth]{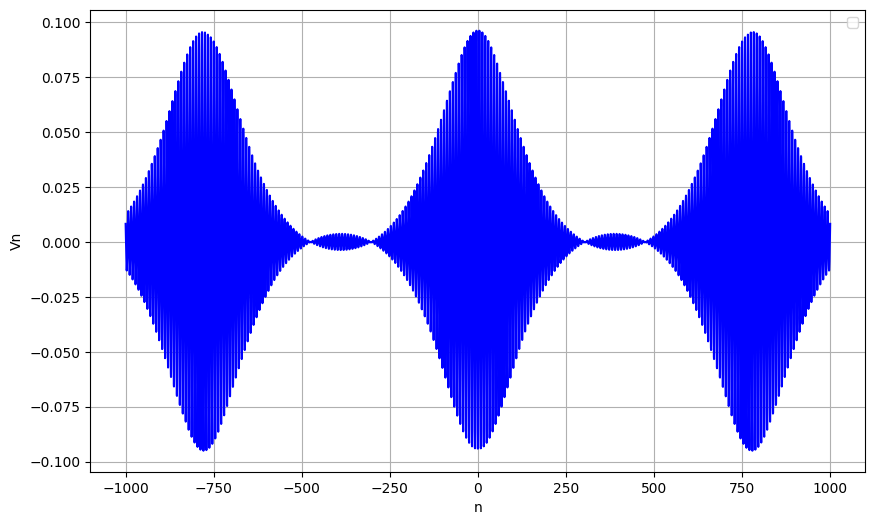}
			\caption{}
		\end{subfigure}
		\begin{subfigure}{0.45\textwidth}
			%\caption{}
			\includegraphics[width=\linewidth]{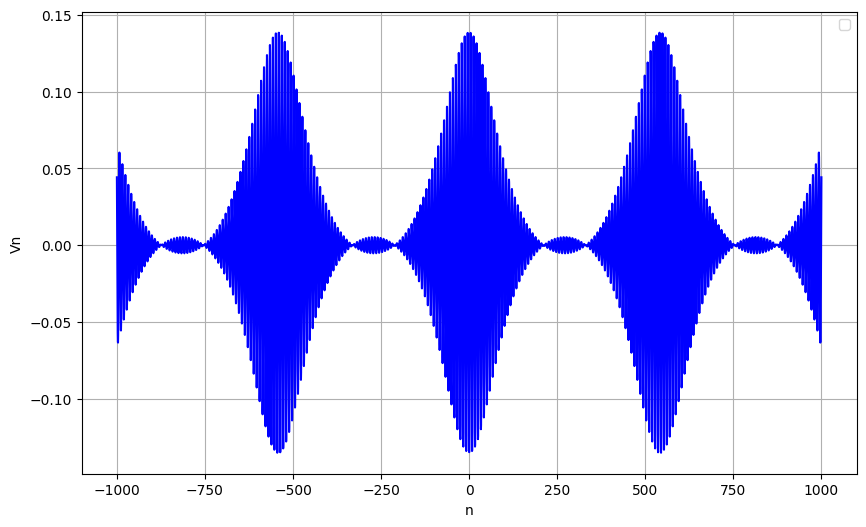}
			\caption{}
		\end{subfigure}
	\end{center}
	\vspace{-0.3cm}
	\caption{ Dynamics of Akhmediev breathers in an electrical transmission line (\ref{eq2}) at $t=0$ and $c=1$ with varying $\rho$: (a) $\rho=0.3$, (b) $\rho=1.3$, (c) $\rho=2.3$, (d) $\rho=3.3$.}
	\label{figb4}
\end{figure*}

In Fig. \ref{figb4}, the dynamics of Akhmediev breathers in an electrical transmission line are presented with the varying parameter $\rho$ at $t=0$. The other parameters are $k=0.75$, $\epsilon=0.01$, $c_0 = 0.0001, d_1 = 0, d_2 = 0$ and $c_0$ is zero. This study is similar to the previous case (Fig. \ref{figb3}) but focuses on different $\rho$ values. In Fig. \ref{figb4}(a), for $\rho=0.3$, the waveform of the breather is displayed. When the $\rho$ value is increased to $1.3$, the shape of the waveform changes, as shown in Fig. \ref{figb4}(b). The amplitude also changes; in Fig. \ref{figb4}(a), it reaches $0.01$, while in Fig. \ref{figb4}(b), it attains $0.04$. Further increasing $\rho$ to $2.3$ and $3.3$ results in an increase in amplitude to $0.1$ and $0.15$, respectively, and the breather form is demonstrated in Figs. \ref{figb4}(c) and \ref{figb4}(d). From Figs. \ref{figb4}(a) and \ref{figb4}(b), there is no breather structure (which means it started to form a breather structure), but in Figs. \ref{figb4}(c) and \ref{figb4}(d), the breather structure is formed, indicating the development of breather peaks. From this, we observed that the waveform of the breather-indicating the development of breather peaks and an increase in amplitude-intensifies as we increase the value of $\rho$  within the range $n= -1000$ to $1000$.

\begin{figure*}[!ht]
	\begin{center}
		\begin{subfigure}{0.45\textwidth}
			%\caption{}
			\includegraphics[width=\linewidth]{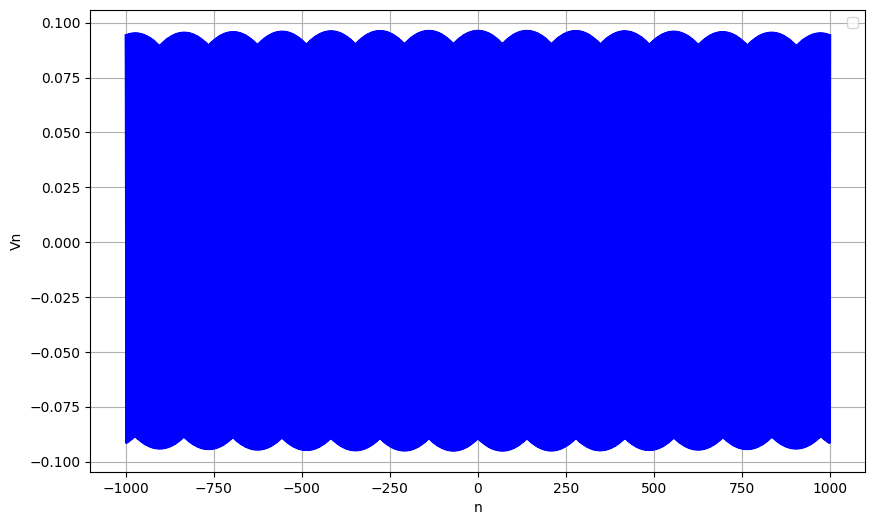}
			\caption{}
		\end{subfigure}
		\begin{subfigure}{0.45\textwidth}
			%\caption{}
			\includegraphics[width=\linewidth]{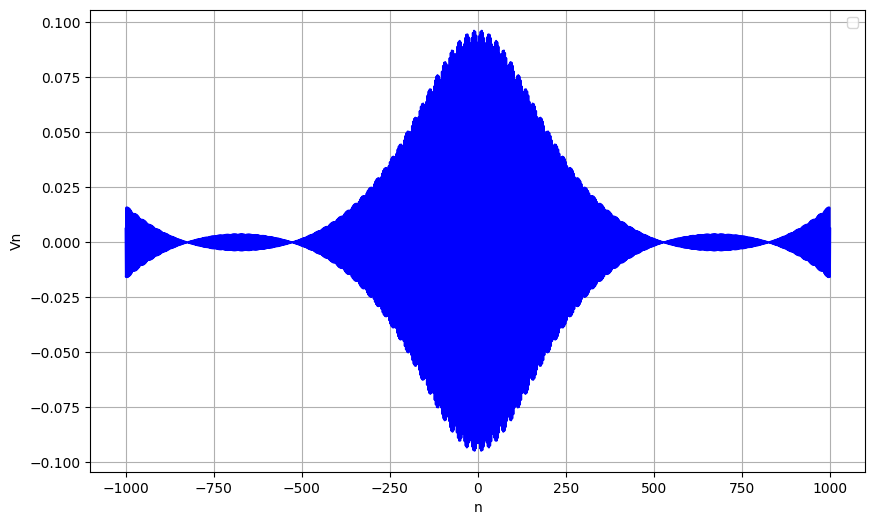}
			\caption{}
		\end{subfigure}\\
		\begin{subfigure}{0.45\textwidth}
			%\caption{}
			\includegraphics[width=\linewidth]{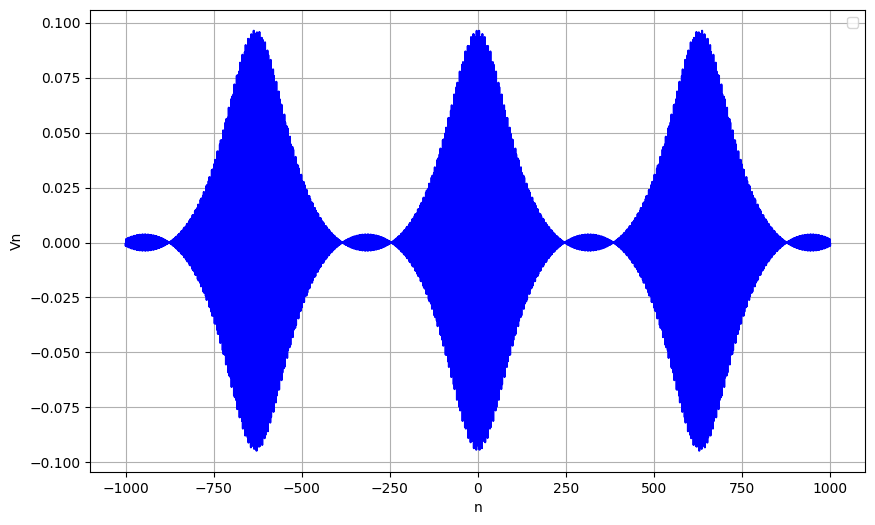}
			\caption{}
		\end{subfigure}
		\begin{subfigure}{0.45\textwidth}
			%\caption{}
			\includegraphics[width=\linewidth]{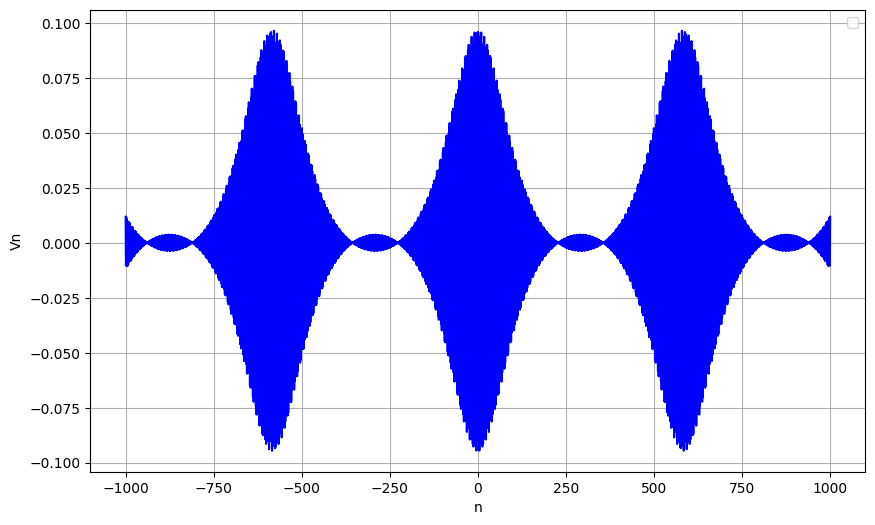}
			\caption{}
		\end{subfigure}
	\end{center}
	\vspace{-0.3cm}
	\caption{Dynamics of Akhmediev breathers in an electrical transmission line (\ref{eq2}) at $t=0$ and $c=1$ for varying $k$ values: (a) $k=1.75$, (b) $k=2$, (c) $k=2.75$ and (d) $k=3$.}
	\label{figb5}
\end{figure*} 

Dynamics of Akhmediev breathers in an electrical transmission line are illustrated in Fig. \ref{figb5} with varying values of $k$ at $t=0$. The other parameters are $\rho=2.3$, $\epsilon=0.01$, $c_0 = 0.0001, d_1 = 0, d_2 = 0$ and $c_1$ is zero. This analysis builds on the previous study (Figs. \ref{figb3} and \ref{figb4}), focusing specifically on different $k$ values. 
When we increase the values of $k$ to $1.75,2, 2.75$ and $3$, the dynamics of the waveform change, but the amplitude remains unchanged, as shown in Figs. \ref{figb5}(a), \ref{figb5}(b), \ref{figb5}(c), and \ref{figb5}(d), respectively. The changes occur within the range of $n$ from $-1000$ to $1000$. From these observations, we conclude that the parameter $k$ is responsible for the increase in the breather peaks and the behavior of the waveform. For $k=1.75$, the peaks of the waveform appear more localized and pronounced, resulting in a darker appearance due to the increased density of points involved. This leads to larger gaps between the waves. Conversely, at $k=3$, the gaps between the waves become smaller, and the localization is reduced, resulting in a smoother breather waveform.
\begin{figure*}[!ht]
	\begin{center}
		\begin{subfigure}{0.45\textwidth}
			%\caption{}
			\includegraphics[width=\linewidth]{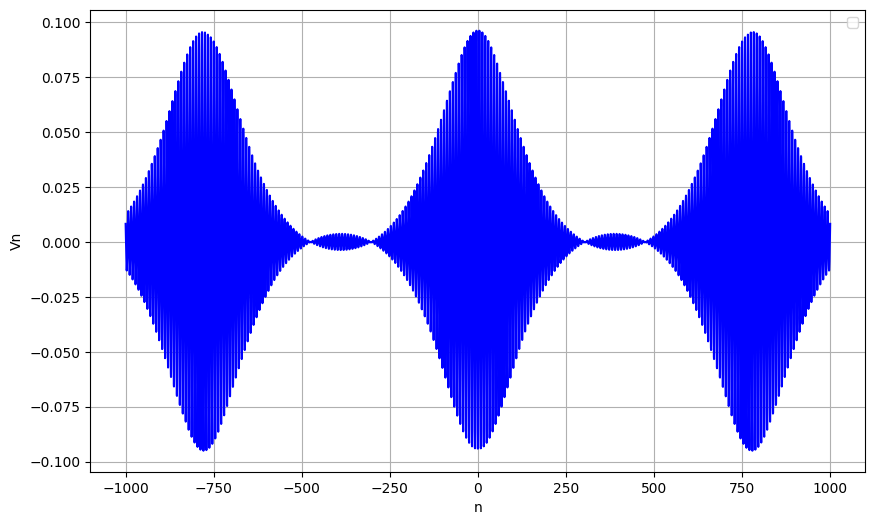}
			\caption{}
		\end{subfigure}
		\begin{subfigure}{0.45\textwidth}
			%\caption{}
			\includegraphics[width=\linewidth]{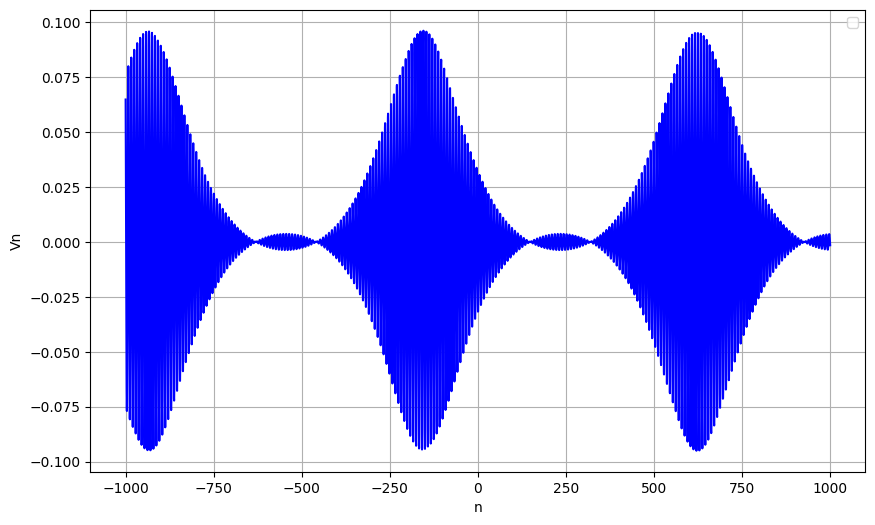}
			\caption{}
		\end{subfigure}\\
		\begin{subfigure}{0.45\textwidth}
			%\caption{}
			\includegraphics[width=\linewidth]{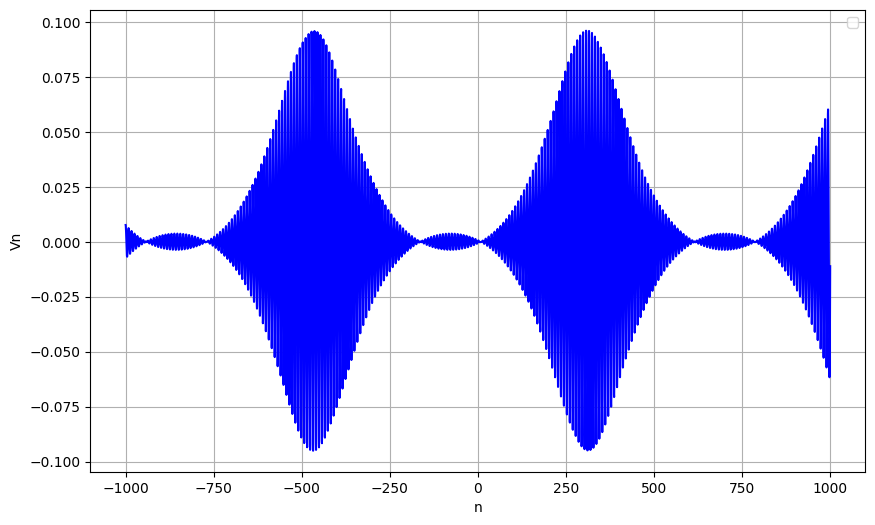}
			\caption{}
		\end{subfigure}
		\begin{subfigure}{0.45\textwidth}
			%\caption{}
			\includegraphics[width=\linewidth]{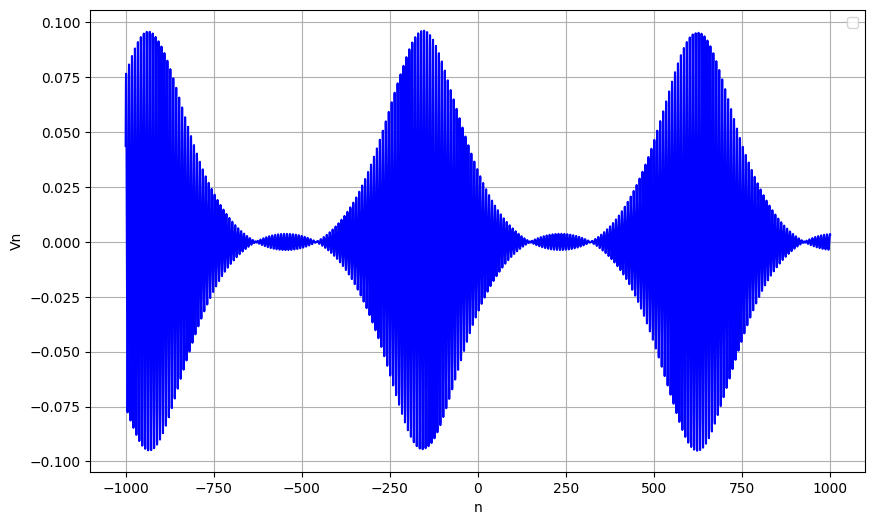}
			\caption{}
		\end{subfigure}
	\end{center}
	\vspace{-0.3cm}
	\caption{Dynamics of Akhmediev breathers in an electrical transmission line (\ref{eq2}) for $c=1$ and different values of $t$: (a) $t=0$, (b) $t=10$, (c) $t=30$ and (d) $t=60$.}
	\label{figb50}
\end{figure*} 

In Fig. \ref{figb50}, we illustrate the dynamics of Akhmediev breathers in the electrical transmission line at different time intervals. The additional parameters are $\rho=2.3$, $k=0.75$, $\epsilon=0.01$, $c_0 = 0.0001, d_1 = 0, d_2 = 0$ and $c_1$ is zero. In Fig. \ref{figb50}(a), within the range $n=-1000$ to $n=1000$, a single peak is observed in the breather and this peak begins to propagate at $t = 0$ and continues to shift forward at $t = 10$, as shown in Fig. \ref{figb50}(b). It is noteworthy that between the peaks, periodic oscillations occur—characterized by small oscillations followed by a large peak—forming a repeating pattern consistent with breather structures reported in the literature. At $t = 30$ and $t = 60$, small periodic oscillations appear between two peaks, as depicted in Fig. \ref{figb50}(c) and  Fig. \ref{figb50}(d) respectively. This indicates that as time progresses, this peak propagates forward, and an additional peak begins to emerge. Interestingly, both the emerging peak and the forward-moving peak appear as partially formed, resembling half peaks rather than fully developed ones. From these observations, we conclude that as time increases, the breather peaks propagate along the transmission line without any change in amplitude. The formation of new peaks occurs gradually, and both the emerging and propagating peaks appear in an incomplete form. This behavior highlights the dynamic evolution of Akhmediev breathers, demonstrating the redistribution of energy and the progressive development of complex oscillatory structures over time.

In the following, we summarize how the parameters $\epsilon$, $\rho$, $k$ and $t$ collectively influence the formation and characteristics of the Akhmediev breather in the modified Noguchi electrical transmission line. Specifically, $\epsilon$ controls the oscillatory intensity and frequency, with larger values leading to stronger oscillations and more complex waveforms. The parameter $\rho$ governs the amplitude and strength of the breather, with higher values resulting in larger, more intense peaks. Meanwhile, $k$ affects the localization and spacing of the peaks, whereas smaller values create broader waves, while larger values increase the number of breather peaks in the same region while maintaining a constant amplitude. Additionally, the parameter $t$ represents the temporal evolution of the breather, influencing the movement of its peaks over time. As $t$ increases, the breather’s peaks shift their positions within the transmission line, while the amplitude remains unchanged. This effect becomes particularly noticeable as $t$ increases, with new peaks appearing, indicating the continuous evolution of the breather. Together, these parameters determine the shape, amplitude, and distribution of the Akhmediev breather, directly impacting the signal quality and performance in the transmission system. By manipulating these parameters, one can effectively analyze and control the formation, behavior, and disappearance of Akhmediev breathers in the transmission line, helping optimize the signal's stability and performance.
\section{Super RW solutions for the electrical transmission line }
\subsection{A super RW solution for the NLS equation (\ref{eq12}) and (\ref{eq9})}
In this section, we derive the super RW solution for Eq. (\ref{eq9}), and then we connect the solutions to the experimental coordinates. Finally, we analyze the dynamics of super RW solution in the electrical transmission line. To do this, we consider the super RW solution of Eq. (\ref{eq12}) in the form \cite{Chabchoub2012super,Duan2020super}
\begin{subequations}
	\label{eq13}
	\begin{equation}
	\Psi_r(X,T)=\left(1-\frac{G+iH}{D}\right)e^{2iT},
	\end{equation} 
	where 
	\begin{align}
	G=&\left(X^2+4T^2 +\frac{3}{4}\right)\left(X^2+20T^2 +\frac{3}{4}\right)-\frac{3}{4},\\
	H=&2T(4T^2-3X^2)+2T\left(2(X^2+4T^2)^2-\frac{15}{8}\right),\\
	D=&\frac{1}{3}(X^2+4T^2)^3+\frac{1}{4}(X^2-12T^2)^2+ \frac{3}{64}(12X^2+76T^2+1).
	\end{align}
\end{subequations}
A notable feature of this solution is its ability to enhance the peak amplitude of the carrier wave at $X=0$ and $T=0$ by a factor of $5$.

Rogue waves, such as those described by the Peregrine soliton, typically amplify the peak wave amplitude by a factor of $3$, making them large and rare occurrences in nonlinear, dispersive media. In contrast, super RWs are characterized by even greater amplification, often reaching a factor of $5$ or more \cite{Duan2020super}. This results in physically larger, steeper, and more extreme waves, caused by higher-order nonlinearities and intensified energy focusing. While both wave types originate from similar mechanisms, super RWs represent a more extreme and dangerous form of RWs.

Using the above solution (\ref{eq13}), we can derive the super RW solution of Eq.(\ref{eq9}) through (\ref{eq11}), which in turn reads
\begin{subequations}
	\label{eq131}
	\begin{equation}
	\Psi_r(\zeta,\tau)=\left(1-\frac{G_1+iH_1}{D_1}\right)e^{i[(\lambda_1\zeta+\lambda_2)+Q\rho^2\tau]},
	\end{equation} 
	with
	\begin{align}
	G_1=&-\frac{3}{4} + \left(\frac{3}{4} + \frac{Q x^2 \rho^2}{2P} + Q^2 \rho^2 \tau^2\right)
	\left(\frac{3}{4} + \frac{Q x^2 \rho^2}{2P} + 5 Q^2 \rho^2 \tau^2\right),\\
	H_1=&Q \rho \tau \left(-\frac{3 Q x^2 \rho^2}{2P} + Q^2 \rho^2 \tau^2\right) 
	+ Q \rho \tau \left(-\frac{15}{8} + 2 \left(\frac{Q x^2 \rho^2}{2P} + Q^2 \rho^2 \tau^2\right)^2\right),\\
	D_1=&\frac{1}{4} \left(\frac{Q x^2 \rho^2}{2P} - 3 Q^2 \rho^2 \tau^2\right)^2 
	+ \frac{1}{3} \left(\frac{Q x^2 \rho^2}{2P} + Q^2 \rho^2 \tau^2\right)^3 
	\nonumber\\&+ \frac{3}{64} \left(1 + \frac{6 Q x^2 \rho^2}{P} + 44 Q^2 \rho^2 \tau^2\right).
	\end{align}
\end{subequations}
The solution (\ref{eq131}) differs from the one presented in Refs. \cite{Kengne2017modelling, Kengne2019transmission,Djelah2023rogue}. Specifically, in the work of Kengne et al., while the same solution is considered, in our formulation we considered $\zeta = \frac{X+VT}{\rho \sqrt{\frac{2P}{Q}}}$, with a real constant $\rho$, whereas in Kengne's paper, the transformation is considered in the form $\zeta = x + X_1(\tau)$ \cite{Kengne2019transmission}. The dynamics, shape, amplitude, and width of the wave which we consider differ significantly from the earlier study, leading to a distinct analytical and physical insights.
\begin{figure*}[!ht]
	\begin{center}
		\begin{subfigure}{0.45\textwidth}
			%\caption{}
			\includegraphics[width=\linewidth]{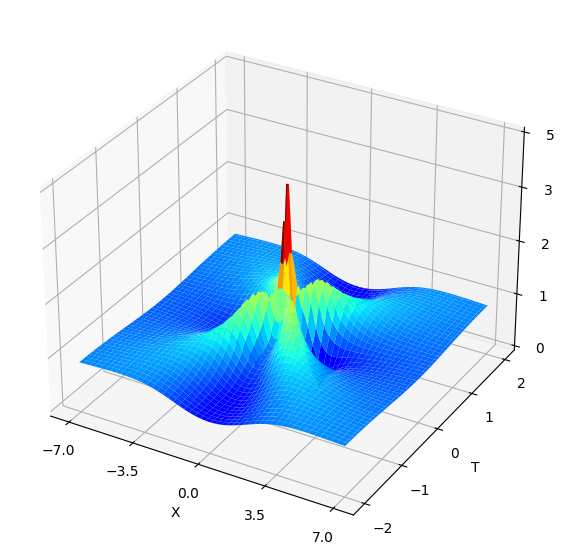}
			\caption{}
		\end{subfigure}
		\begin{subfigure}{0.45\textwidth}
			%\caption{}
			\includegraphics[width=\linewidth]{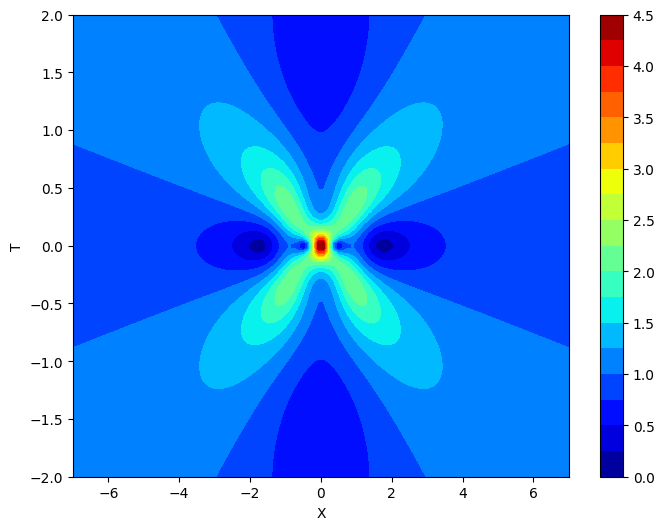}
			\caption{}
		\end{subfigure}
	\end{center}
	\vspace{-0.3cm}
	\caption{Super RW ($|\Psi_r(X,T)|$) for the NLS equation (\ref{eq12}). (a) 3D plot and (b) contour plot.}
	\label{fig3}
\end{figure*} 

The super RW for the NLS equation (\ref{eq12}) is shown in Fig. \ref{fig3}. Figures \ref{fig3}(a) and \ref{fig3}(b) present the 3D plot and contour plot of the super RW for the NLS equation (\ref{eq12}), respectively. From these figures, we can observe that the amplitude of the rogue wave is larger than that of the Peregrine soliton, as shown in Ref. \cite{Duan2020super}. The amplitude of the super RW attains its maximum level of $4.5$ (approximately) at $(0,0)$. 

\begin{figure*}[!ht]
	\begin{center}
		\begin{subfigure}{0.45\textwidth}
			%\caption{}
			\includegraphics[width=\linewidth]{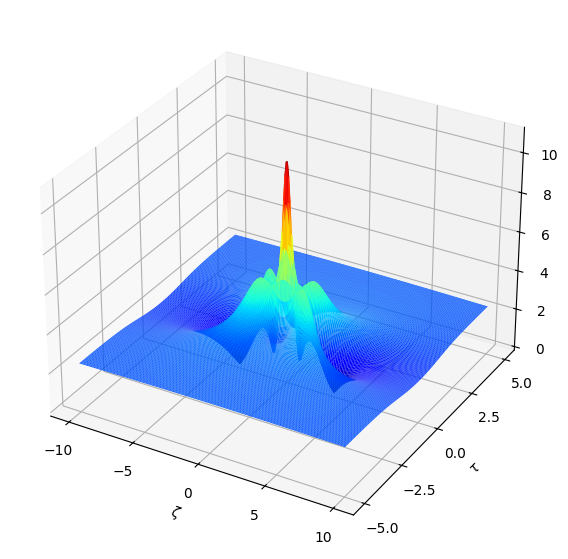}
			\caption{}
		\end{subfigure}
		\begin{subfigure}{0.45\textwidth}
			%\caption{}
			\includegraphics[width=\linewidth]{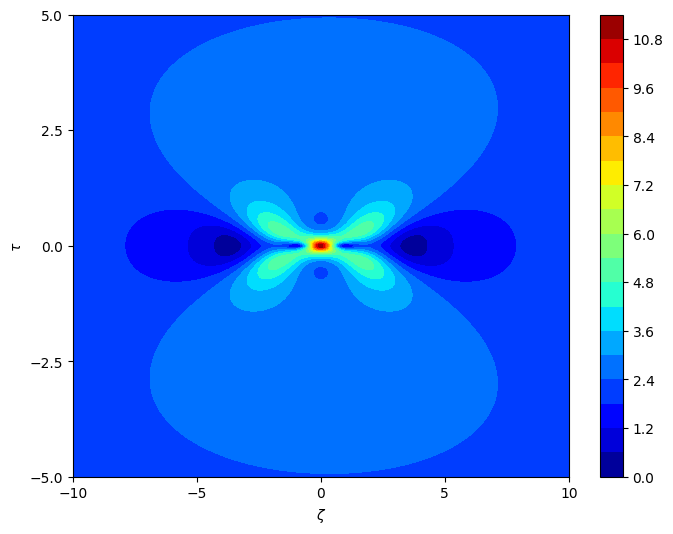}
			\caption{}
		\end{subfigure}\\\begin{subfigure}{0.45\textwidth}
			%\caption{}
			\includegraphics[width=\linewidth]{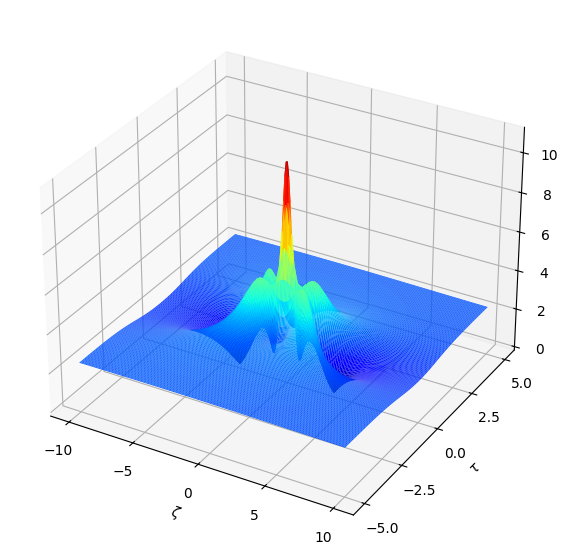}
			\caption{}
		\end{subfigure}
		\begin{subfigure}{0.45\textwidth}
			%\caption{}
			\includegraphics[width=\linewidth]{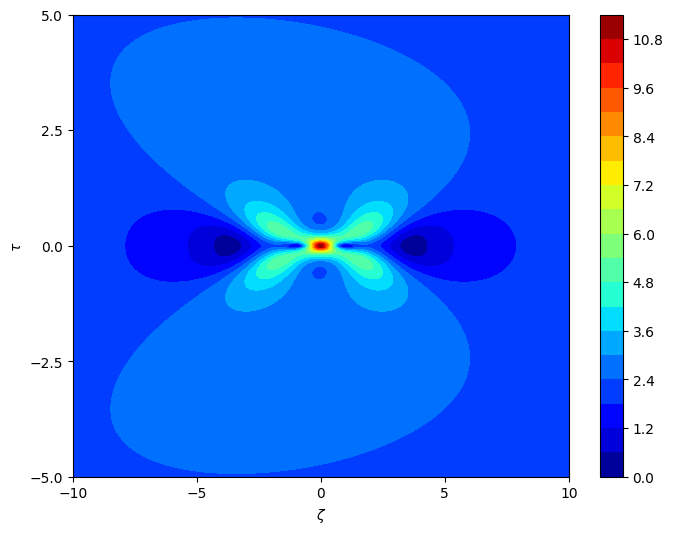}
			\caption{}
		\end{subfigure}
	\end{center}
	\vspace{-0.3cm}
	\caption{Super RW ($|\psi_r(\zeta,\tau)|$) for the NLS equation (\ref{eq9}). (a), (b): 3D and contour plots for $c_0=0.0001$, (c), (d): 3D and contour plots for $c_0=0.001$.}
	\label{fig4}
\end{figure*}

Figure \ref{fig4} presents the same data as Figure \ref{fig3} but in experimental coordinates, depicting the solution of Eq. (\ref{eq11}) with $c_0 = 0.0001$. In these experimental coordinates, the super RW appears over the $\zeta$ and $\tau$ axis ranges from $-10$ to $10$ and from $-5$ to $5$, respectively. The maximum amplitude of the super RW in this setting is approximately $10.8$, and its width also varies, as shown in the 3D and contour plots in Figures \ref{fig4}(a) and \ref{fig4}(b). Figures \ref{fig4}(c) and \ref{fig4}(d) correspond to Figures \ref{fig4}(a) and \ref{fig4}(b) but for $c_0 = 0.001$. Here, we observe a loss in the super RW, as shown in Figures \ref{fig4}(c) and \ref{fig4}(d). We also verify that the RWs experience greater loss for higher values of $c_0$.
\subsection{A super RW solution in the modified Noguchi electrical transmission line (\ref{eq2})}
To construct the super RW solution of the considered electrical transmission line, we consider
\begin{equation}
V_{nr}(t)=\epsilon\psi_r(\zeta,\tau)e^{i\theta}+\epsilon^2\psi_{10}(\zeta,\tau)+\epsilon^2 \psi_{20}(\zeta,\tau)e^{2i\theta}+c.c.,
\label{eq14}
\end{equation}
where $\psi_r(\zeta,\tau)$, $\psi_{10}(\zeta,\tau)$ and $\psi_{20}(\zeta,\tau)$ are given in Eqs. (\ref{eq131}), (\ref{eq8}) and (\ref{eq81}). Equation (\ref{eq14}) describes the dynamics of super RWs in an electrical transmission line.

\begin{figure*}[!ht]
	\begin{center}
		\begin{subfigure}{0.45\textwidth}
			%\caption{}
			\includegraphics[width=\linewidth]{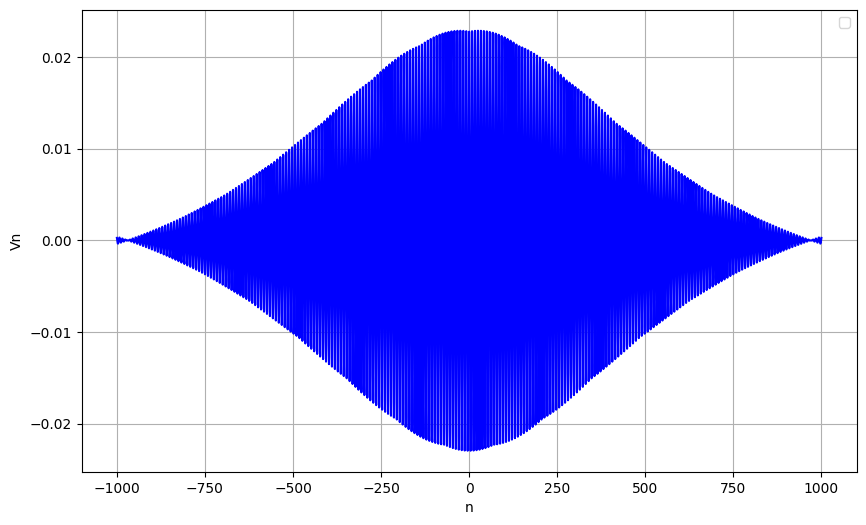}
			\caption{}
		\end{subfigure}
		\begin{subfigure}{0.45\textwidth}
			%\caption{}
			\includegraphics[width=\linewidth]{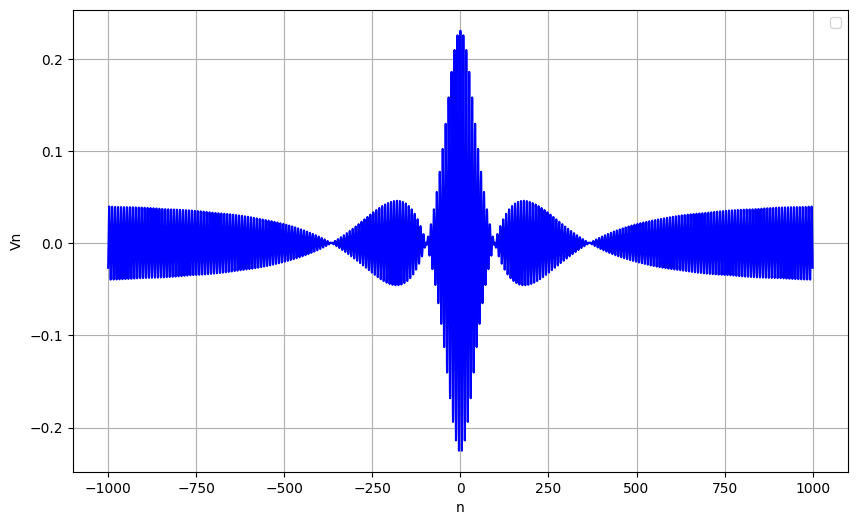}
			\caption{}
		\end{subfigure}\\
		\begin{subfigure}{0.45\textwidth}
			%\caption{}
			\includegraphics[width=\linewidth]{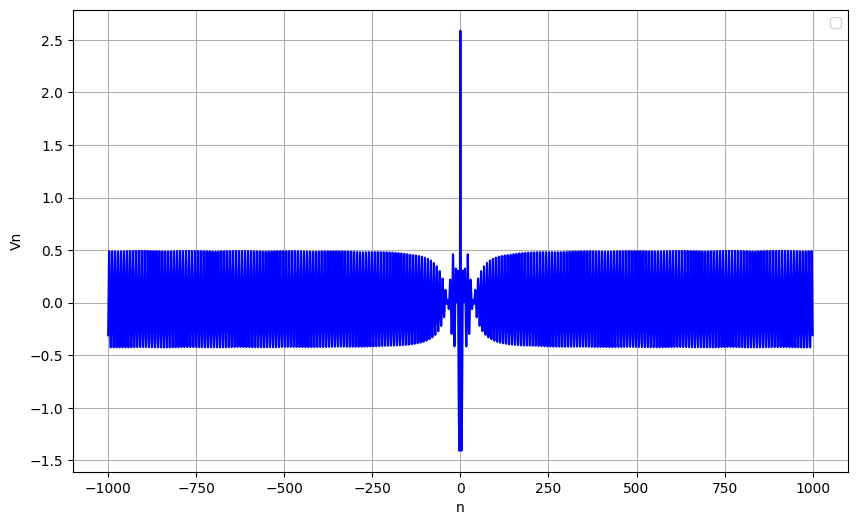}
			\caption{}
		\end{subfigure}
		\begin{subfigure}{0.45\textwidth}
			%\caption{}
			\includegraphics[width=\linewidth]{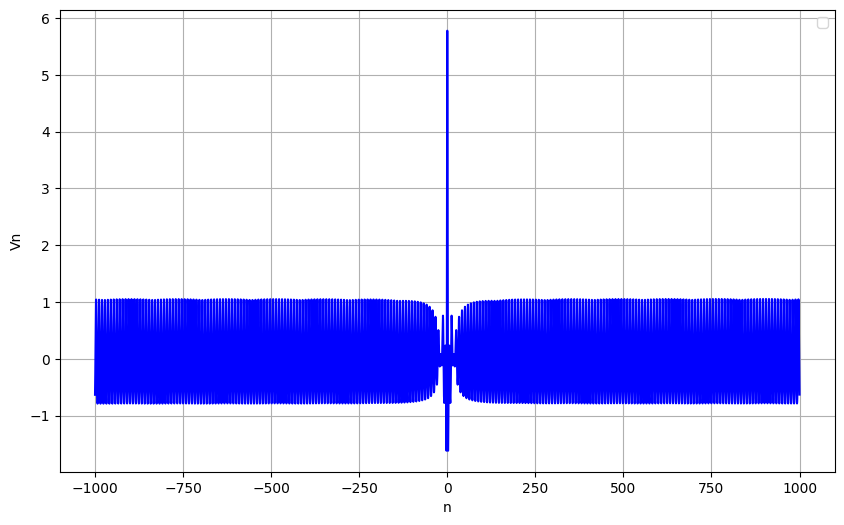}
			\caption{}
		\end{subfigure}
	\end{center}
	\vspace{-0.3cm}
	\caption{Variation of super RW in an electrical transmission line (\ref{eq2}) with parameter $\epsilon$ at $t=0$. (a) $\epsilon=0.001$; (b) $\epsilon=0.01$; (c) $\epsilon=0.1$; (d) $\epsilon=0.2$.}
	\label{fig5}
\end{figure*} 

Figure \ref{fig5} illustrates the super RW in the electrical transmission line, as described by Eq. (\ref{eq14}), with different values of the parameter $\epsilon$ at $t=0$. The parameters $k=0.75$, $\rho=2.3$, and $c_0 = 0.0001$ are fixed, while $d_1$, $d_2$ and $c_1$ are set to zero. Adjustments in $\epsilon$ modify the shape and amplitude of the super RW, showing how $\epsilon$ influences the wave profile under otherwise constant conditions. In Fig. \ref{fig5}(a), with $\epsilon=0.001$, the initial formation of the super RW appears, though it is not fully developed. As $\epsilon$ increases to $0.01$, the RW reaches its full form, as seen in Figure \ref{fig5}(b). Also, the amplitude of the super RW is obtained at $0.2$. A further increase in $\epsilon$ to $0.1$ and $0.2$ causes a rise in amplitude (from $2.5$ to $6$) and an increase in the number of background waves, as illustrated in Figs. \ref{fig5}(c) and \ref{fig5}(d). These observations indicate that as $\epsilon$ increases, the RW shape becomes more complex, higher amplitude, and increased oscillations in the background waves. This sensitivity to $\epsilon$ highlights its key role in determining the RW's behavior, with implications on signal quality and transmission system performance.

\begin{figure*}[!ht]
	\begin{center}
		\begin{subfigure}{0.45\textwidth}
			%\caption{}
			\includegraphics[width=\linewidth]{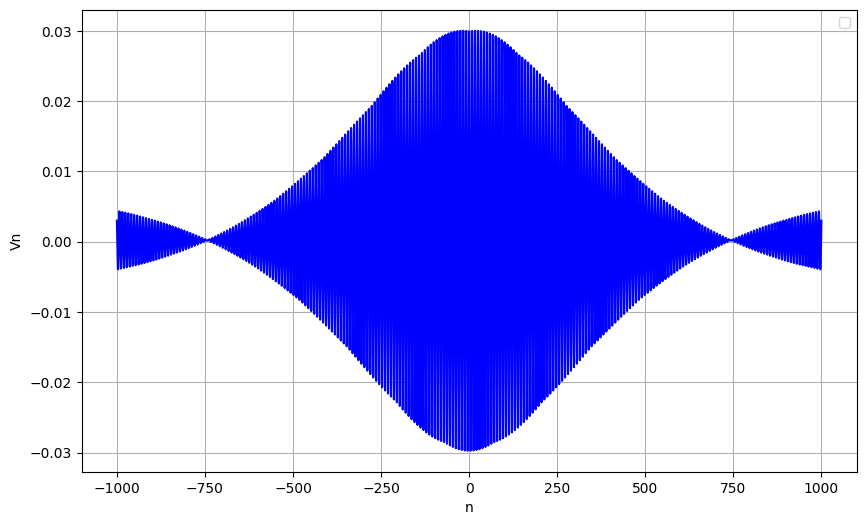}
			\caption{}
		\end{subfigure}
		\begin{subfigure}{0.45\textwidth}
			%\caption{}
			\includegraphics[width=\linewidth]{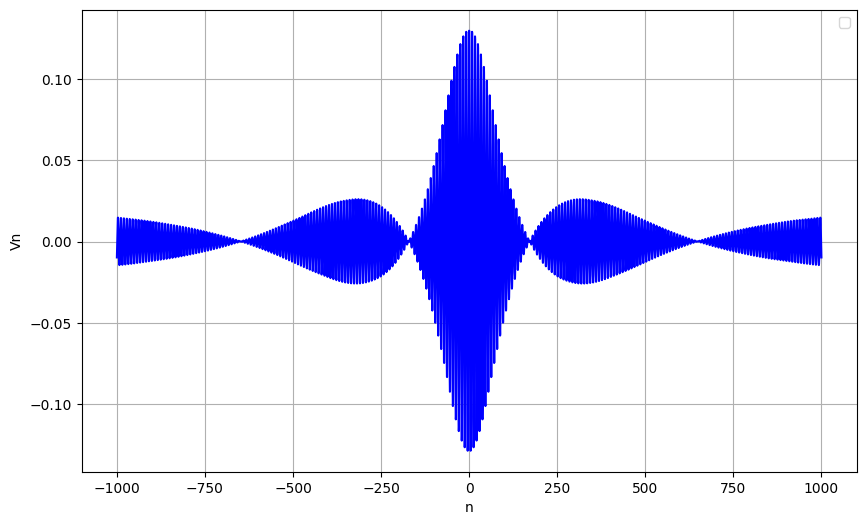}
			\caption{}
		\end{subfigure}\\
		\begin{subfigure}{0.45\textwidth}
			%\caption{}
			\includegraphics[width=\linewidth]{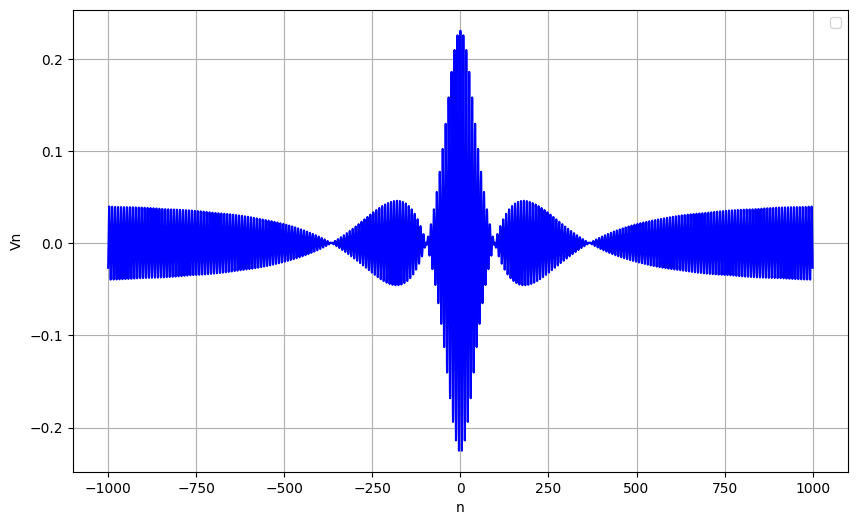}
			\caption{}
		\end{subfigure}
		\begin{subfigure}{0.45\textwidth}
			%\caption{}
			\includegraphics[width=\linewidth]{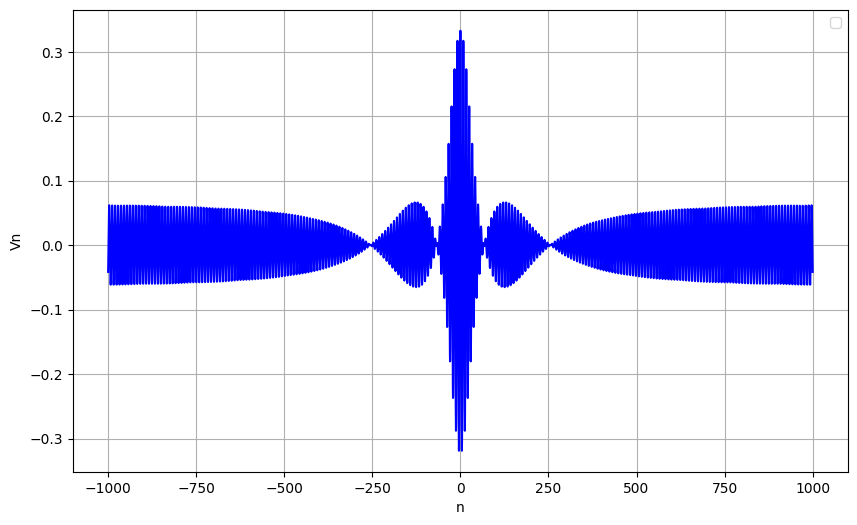}
			\caption{}
		\end{subfigure}
	\end{center}
	\vspace{-0.3cm}
	\caption{Dynamics of super RW in an electrical transmission line (\ref{eq2}) at $t=0$ with varying $\rho$: (a) $\rho=0.3$, (b) $\rho=1.3$, (c) $\rho=2.3$, (d) $\rho=3.3$.}
	\label{fig6}
\end{figure*} 

In Fig. \ref{fig6}, the dynamics of super RW in an electrical transmission line are illustrated, focusing on the effects of varying the parameter $\rho$ at $t=0$. The remaining parameters are maintained at $k=0.75$, $\epsilon=0.01$, $c_0 = 0.0001$, $d_1 = 0$, $d_2 = 0$ and $c_1= 0$. This study parallels the previous case presented in Fig. \ref{figb5}, but it explores different values of $\rho$ to examine how they impact the formation and characteristics of the super RW. In Fig. \ref{fig6}(a), the waveform of the super RW is depicted for $\rho= 0.3$, showing a relatively simple structure. It exhibits a broader oscillation envelope, with denser and more spread-out oscillations extending further along the $n$-axis. When the $\rho$ value is increased to $1.3$, as seen in Fig. \ref{fig6}(b), a noticeable change in the shape of the waveform occurs, indicating a super RW profile. The amplitude also shifts significantly; in Fig. \ref{fig6}(a), the amplitude reaches $0.03$, while in Fig. \ref{fig6}(b), it increases to $0.1$. As $\rho$ continues to increase to $2.3$ and $3.3$, the amplitude further rises to $0.2$ and $0.3$, respectively, as shown in Figs. \ref{fig6}(c) and \ref{fig6}(d). Also, the width of the super RW decreases (which means super RW becomes sharper), and the number of background waves increases, indicating a change in the overall wave profile. This progress illustrates how the variations in $\rho$ play a crucial role in shaping the characteristics and behavior of super RWs in electrical transmission lines.

\begin{figure*}[!ht]
	\begin{center}
		\begin{subfigure}{0.45\textwidth}
			%\caption{}
			\includegraphics[width=\linewidth]{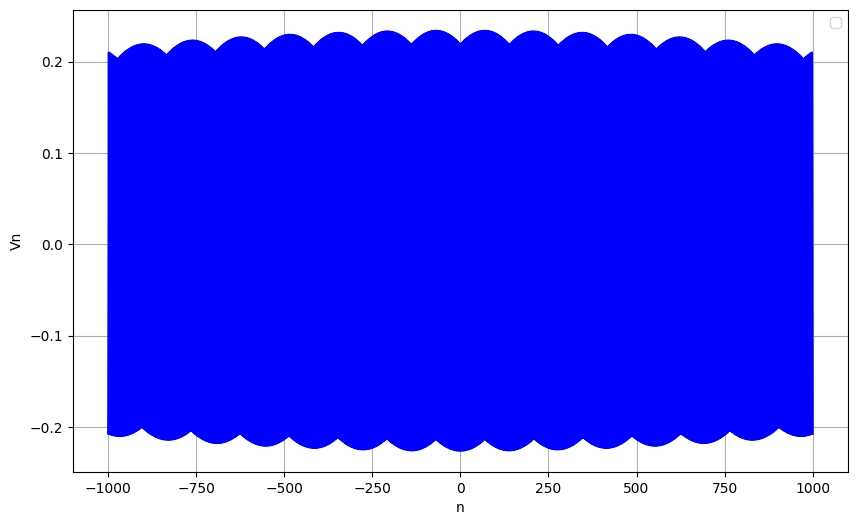}
			\caption{}
		\end{subfigure}
		\begin{subfigure}{0.45\textwidth}
			%\caption{}
			\includegraphics[width=\linewidth]{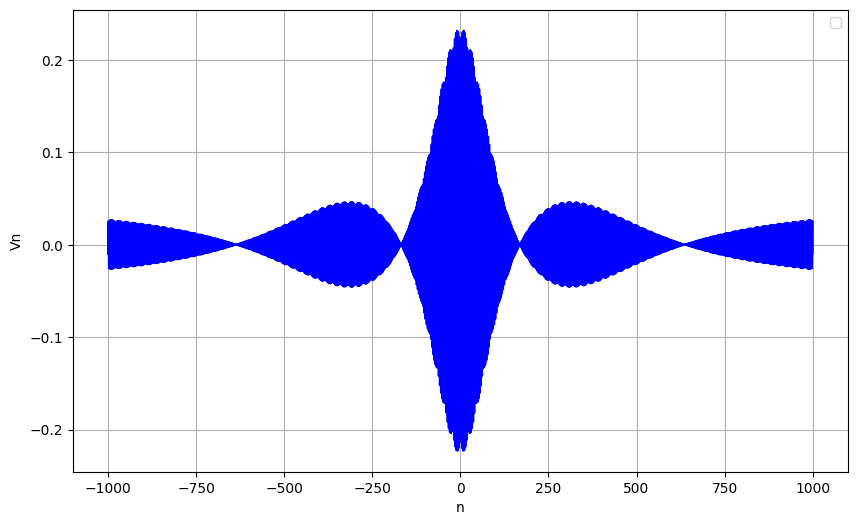}
			\caption{}
		\end{subfigure}\\
		\begin{subfigure}{0.45\textwidth}
			%\caption{}
			\includegraphics[width=\linewidth]{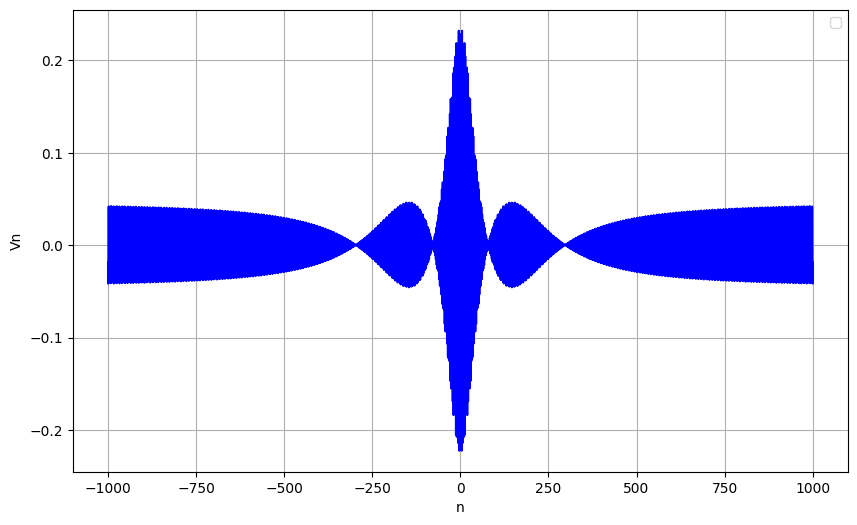}
			\caption{}
		\end{subfigure}
		\begin{subfigure}{0.45\textwidth}
			%\caption{}
			\includegraphics[width=\linewidth]{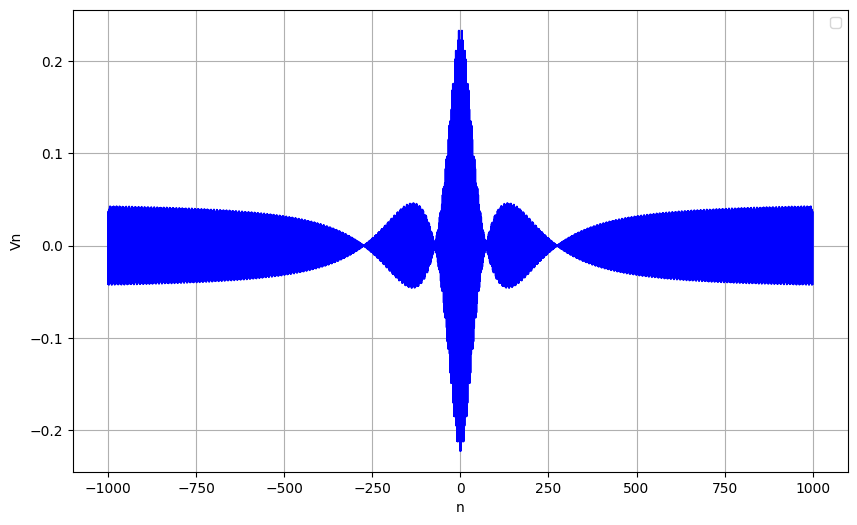}
			\caption{}
		\end{subfigure}
	\end{center}
	\vspace{-0.3cm}
	\caption{Dynamics of super RW in an electrical transmission line (\ref{eq2}) at $t=0$ for varying $k$ values: (a) $k=1.75$, (b) $k=2$, (c) $k=2.75$ and (d) $k=3$.}
	\label{fig7}
\end{figure*} 
The dynamics of super RW in the electrical transmission line are illustrated in Figure \ref{fig7}, showcasing varying values of $k$ at $t=0$. The other parameters are set to $\rho=2.3$, $\epsilon=0.01$, $c_0 = 0.0001, d_1 = 0$, $d_2 = 0$ and $c_1= 0$. This analysis builds on the previous studies presented in Figures \ref{fig5} and \ref{fig6}, focusing specifically on different $k$ values. As we increase the values of $k$ to $1.75,2, 2.75$ and $3$, the dynamics of the super RW waveform change, especially there is a change in the shape of the waves, as shown in Figures \ref{fig7}(a), \ref{fig7}(b), \ref{fig7}(c), and \ref{fig7}(d), respectively. In all the cases (\ref{fig7}(a), \ref{fig7}(b), \ref{fig7}(c), and \ref{fig7}(d)) the amplitude remain same which is $0.02$. The changes occur within the range of $n$ from $-1000$ to $1000$. Here, the value of $k$ plays a similar role as in the breather case in Fig. \ref{figb5}. The super RW peaks and background waves are more localized for low values of $k$, while higher values lead to less localization. At that time, an explicit super RW can be seen, as shown in Fig. \ref{fig7}(d). In all these figures (Figs. \ref{fig7}(a)-\ref{fig7}(d)), we observe that the width of the super RW and the background waves decreases. These changes occur within the range of $n$ from $-1000$ to $1000$. From these observations, we conclude that the parameter $k$ significantly influences the behavior of the waveform, specifically by changing the shape of the waves and increasing the number background waves as the $k$ value is raised.

\begin{figure*}[!ht]
	\begin{center}
		\begin{subfigure}{0.45\textwidth}
			%\caption{}
			\includegraphics[width=\linewidth]{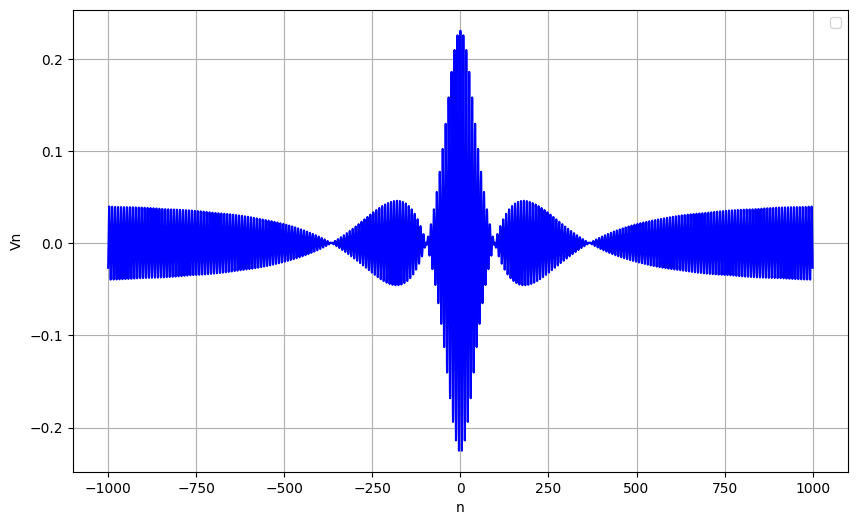}
			\caption{}
		\end{subfigure}
		\begin{subfigure}{0.45\textwidth}
			%\caption{}
			\includegraphics[width=\linewidth]{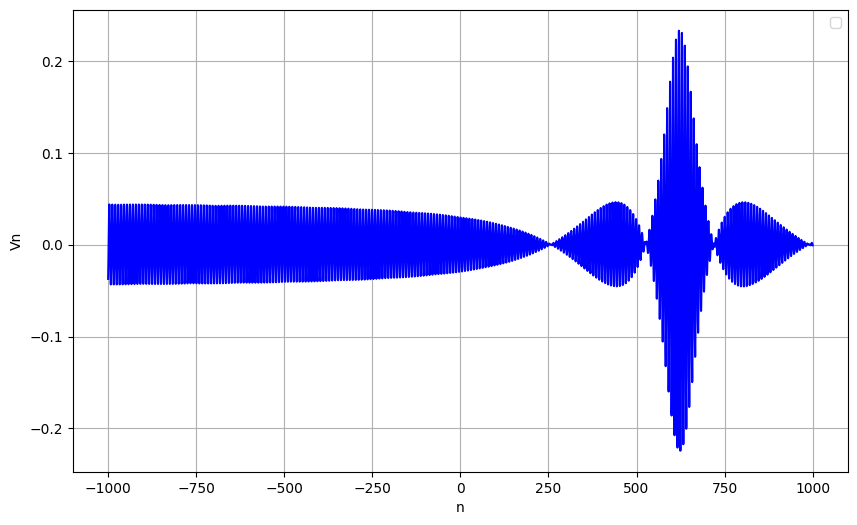}
			\caption{}
		\end{subfigure}\\
		\begin{subfigure}{0.45\textwidth}
			%\caption{}
			\includegraphics[width=\linewidth]{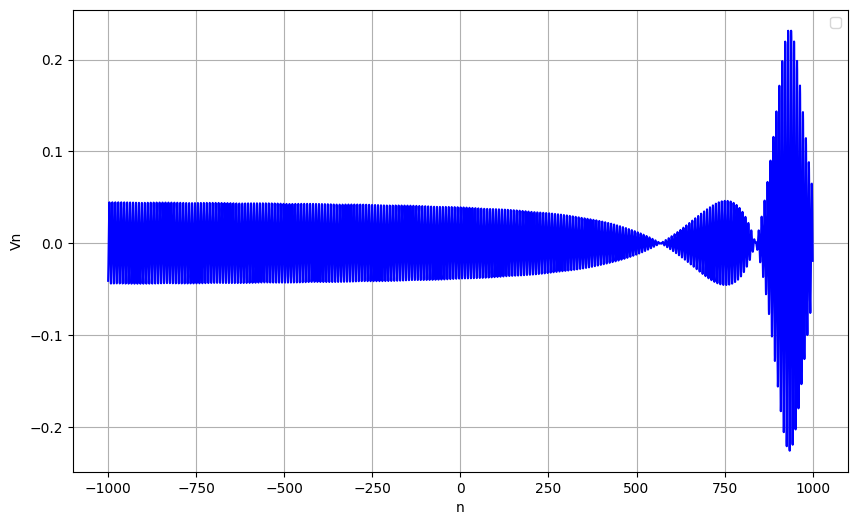}
			\caption{}
		\end{subfigure}
		\begin{subfigure}{0.45\textwidth}
			%\caption{}
			\includegraphics[width=\linewidth]{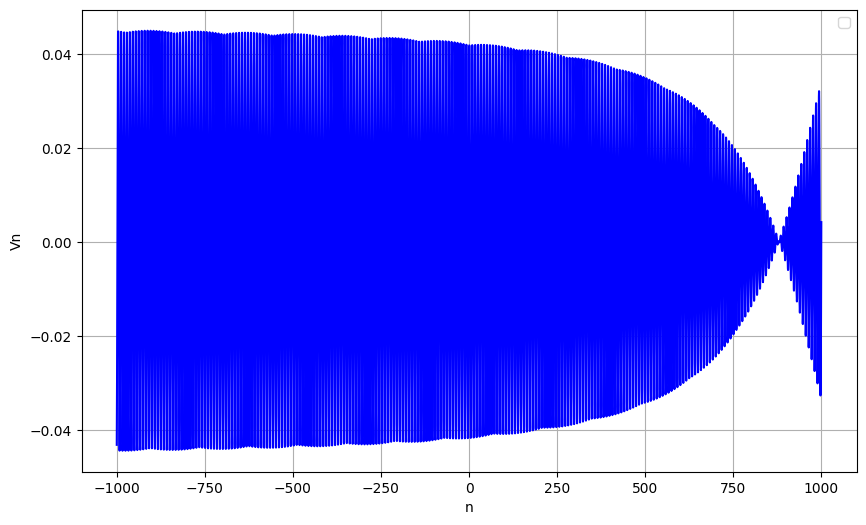}
			\caption{}
		\end{subfigure}
	\end{center}
	\vspace{-0.3cm}
	\caption{Dynamics of super RW in an electrical transmission line (\ref{eq2}) for different values of time: (a) $t=0$, (b) $t=10$, (c) $t=15$ and (d) $t=20$.}
	\label{fig71}
\end{figure*} 

The dynamics of the super RW in the electrical transmission line (\ref{eq2}) is illustrated in Fig. \ref{fig71} for different time values. The remaining parameter values are the same as in Fig. \ref{fig5} ($k = 0.75$, $\rho = 2.3$, $\epsilon=0.01$ and $c_0 = 0.0001$ are kept fixed, while $d_1$, $d_2$, and $c_1$ are set to zero.). For $t = 0$, the dynamics of the super RW in the electrical transmission line is shown in Fig. \ref{fig71}(a). As time increases to $t = 10$, the super RW moves forward. With further increases in time to $t = 15$ and $t = 20$, the super RW moves even further forward, as shown in Figs. \ref{fig71}(b), \ref{fig71}(c), and \ref{fig71}(d), respectively. We observe that the super RW propagates through the electrical transmission line as time increases. Additionally, we note that the amplitude remains the same at $0.2$ for the first three time values, whereas at $t = 20$, the amplitude reduces to $0.04$. This indicates that the background waves are only localized in the $n$ region from $-1000$ to $1000$, while the RW shifts to a new region beyond $n = 1000$, which may reach $n = 1500$ or further. From this, we conclude that the super RW propagates through the electrical transmission line as time increases, with its position shifting forward while its amplitude remains unchanged at $0.2$ for the first three time values and then transitions to $0.04$ at $t = 20$.

Figures \ref{fig5}-\ref{fig71}, reveal that the parameters $\epsilon$, $\rho$ and $k$ play crucial roles in determining the shape, amplitude, and distribution of the super RW in the electrical transmission line. Specifically, $\epsilon$ controls the oscillatory intensity and frequency of the super RW, with larger values leading to stronger oscillations, increased complexity in the waveform, and higher amplitude. $\rho$ governs the amplitude and sharpness of the RW, with higher values resulting in larger, more intense peaks and a decrease in the width of the RW profile. Meanwhile, $k$ influences the localization and spacing of the RW peaks; lower values of $k$ produce more localized peaks, leading to a denser wave, while higher values spread the peaks out, reducing the intensity and background wave presence. In all these figures (Figs. \ref{fig7}(a)-\ref{fig7}(d)), we observe that the width of the super RW and the background waves increases. Additionally, the time parameter $t$ significantly affects the propagation and evolution of the super RW. As time increases, the super RW moves forward along the transmission line, accompanied by a gradual decrease in amplitude and an increase in the width of the RW and background waves.  By manipulating these parameters, one can effectively analyze and control the formation, behavior, and disappearance of super RWs in the transmission line, optimizing signal quality and performance.
\section{Role of Additional Parameters in $V_{nb}$ (Breathers) and $V_{nr}$ (super RW)}
In the above investigations we fixed the parameters, $L_1, L_2, C_0, C_S, \alpha$ and $\beta$. Here, $L_1, L_2$ are denoted in $ mH$, $C_0, C_S,$ are represented in $\mu F$ and, $ \alpha$ and $\beta$ are in $V^{-1}$ and $V^{-2}$, respectively. Now, we analyze the effect of these parameters and examine their influence when they are varied. For breather solution, when we increase the value of $L_1$ from $240$ to $257$, the breather peaks move outward from the origin, and eventually, we observe a breather structure (with less number of breather peaks) in the spatial range $n=-1000$ to $1000$. For the inductance $L_2$, the behavior occurs in reverse: at $L_2=550$, the peaks initially form a less breather peaks structure, but as $L_2$ increases from $550$, the wave moves inward from both sides, toward the center of the origin, and forms a more number of breather peaks structure when it reaches $660$ within the same range. We observe that when we increase $C_0$ from $280$ to $700$ in the range $n= -1000$ to $1000$, initially, the breather solution has more number of peaks with periodic oscillations. However, as $C_0$ is further increased, we observe that only the rogue-like peak remains within the range, with some deviations in the peak structure. For the $C_S$ case, the behavior occurs in reverse: the single rogue-like peak changes to a breather structure, as the value of $C_S$ increases from $36$ to $76$. When considering the value of $\alpha$, for $\alpha = 0.21$, fewer breather peaks are observed. As $\alpha$ decreases to $0.021$, a greater number of breather peaks appear. In the case of $\beta$, for the lower value $0.0197$, fewer breather peaks are present in the breather structure, whereas for the higher value $0.1197$, more breather peaks are formed within the same region $n$.

For super RW solution, when increasing $L_1$ from $160$ to $250$, the dynamics, shape, and size of the super RW increase. Similarly, when increasing $L_2$ from $440$ to $650$, the shape and size of the super RW decrease. When $C_0$ is increased from $280$ to $700$, the shape and size of the super RW in the transmission line become larger and more elaborate. The width of both the RW and the background wave increases, indicating that the wave is not compressed but rather expands outward from the origin. In contrast, when $C_S$ is increased from $16$ to $76$, the width of the super RW decreases, and the wave becomes sharper. The background waves are also compressed toward the origin, showing that the entire wave structure becomes more localized and concentrated. When $\alpha$ is decreased from $0.21$ to $0.021$, the super RW becomes sharper. For $\alpha = 0.021$, both the wave and the background are more localized, whereas for $\alpha = 0.21$, the super RW appears with less localization. Similarly, when $\beta$ is increased, at $0.11$, the super RW and background are more localized and broader, whereas at $1.21$, the super RW becomes more sharper. Here, all the result are analyzed between the $n$ range from $-1000$ to $1000$. All these parameters are also used to analyze the dynamics of both the Akhmediev breather and super RWs in the considered electrical transmission model. 
\section{Conclusion}
In this study, we explored breather and super RW within a modified Noguchi electrical transmission line model. Starting with Kirchhoff's law, we derived a circuit equation and reduced it to a NLS equation with a potential term using the reductive perturbation method. A transformation then allowed us to simplify this further to the standard NLS equation. We examined the dispersion parameter $P$ and nonlinearity parameter $Q$ to investigate the baseband instability region and derived breather and super RW solutions. It is important to emphasize that both breather and super RW observed in our study arise as a result of baseband MI. Mapping these solutions back to the original circuit equation provided insights into wave dynamics within the electrical network. We also analyzed how system parameters shape the behavior of both breathers and super RWs, highlighting their influence on wave propagation in nonlinear systems. In the following we summarize our findings.\\
\textbf{For Breathes:}
(i) Effect of the small parameter, $\epsilon$: Higher $\epsilon$ values make the breather waveform more complex, with increased amplitude and oscillations, suggesting a substantial impact on signal stability. The orientation of the breather remain unchanged. (ii) Role of the real constant, $\rho$: Low $\rho$ values yield a broad, spread-out waveform. As $\rho$ increases (e.g., to $2.3$ and $3.3$), distinct breather peaks emerge, with rising amplitude, indicating that higher $\rho$ values are essential for strong breather profiles. (iii) Influence of the wavenumber, $k$: Lower $k$ values (e.g., $k=1.75$) create broad, localized peaks. Higher $k$ values (e.g., $k=3$) lead to a more spread-in waveform (which means width of the wave becomes decreased) with highly pronounced peaks and a constant amplitude. (iv) Impact of time, $t$: The parameter $t$ represents the temporal evolution of the breather, affecting the movement of its peaks while keeping the amplitude unchanged. As $t$ increases, the peaks shift forward. Also, new peaks appear and move toward the origin, demonstrating the continuous evolution of the breather.\\
\textbf{For Super RWs:}
(i) Effect of the small parameter, $\epsilon$: Increasing $\epsilon$ makes the super RW shape more complex (becomes sharper), with greater amplitude and oscillations in the background, highlighting its role in shaping super RW intensity, which may impact system stability. Additionally, the amplitude of the background waves varies. (ii) Role of the real constant, $\rho$: Rising $\rho$ values (from $0.3$ to $3.3$) result in larger amplitudes and a narrower, more intense super RW profile, increasing the concentration of wave energy and making it more disruptive in transmission systems. (iii) Influence of the wavenumber, $k$: The role of $k$ in the electrical transmission line dynamics is to control the localization and width of the super RW peaks and background waves. Specifically, as $k$ increases, the amplitude of the super RW remains same, while the super RW get more sharper means compression happen strongly for higher values. This means that for lower $k$ values, the super rogue wave is more confined and broader (less sharp), whereas higher $k$ values lead to sharper more localized waveforms. (iv) Impact of time, $t$: The time parameter $t$ significantly influences the propagation and localization of the super RW. As $t$ increases, the super RW moves forward along the transmission line with its amplitude remaining unchanged. For higher values of $t$, the super RW crosses the region ahead of the background waves, which stay confined in that region.

In conclusion, the breather and super RW in our study arise from baseband MI. Notably, we are the first to discuss the concept of baseband MI for the arise of breather and super RW in the context of electrical transmission lines. Additionally, we present breather and super RW solutions in a modified Noguchi electrical transmission line for the first time, particularly emphasizing the breather solution. Furthermore, we analyzed how network parameters $\epsilon$, $\rho$, and $k$ play a critical role in controlling the structure and intensity of both breather and super RW, offering potential approaches for managing waveforms and their effects in nonlinear electrical systems. We have also analyzed both breather and super RW with the network parameters $L_1, L_2, C_0, C_S, \alpha$ and $\beta$ which play a similar role to $\epsilon$, $\rho$, and $k$ in shaping and controlling the waveform characteristics. Finally, we highlight few relevant experimental studies on electrical transmission lines. Aziz et al. proposed a model for solitons in nonlinear transmission lines, validated experimentally, showing solitons with a $\sech^4$ shape \cite{Aziz2020analytical}. Kofane et al. studied pulse solitons in bi-inductance lines, finding that their system reduces to the Korteweg-de Vries equation, with stable propagation for low-amplitude pulses and oscillations for higher-amplitude pulses \cite{Kofane1988theoretical}. Their work also confirmed nonlinear modulated waves in different frequency bands, matching theoretical predictions, including damping effects \cite{Marquie1994generation}. Sekulic et al. focused on solitons in discrete nonlinear transmission lines, showing the impact of losses on wave propagation, with experimental results aligning with theoretical models \cite{Sekulic2021soliton}. Based on these experimental studies and our theoretical findings, our study will help to analyze where breathers and super RWs are likely to arise in experimental setups, providing insights for better wave control. Additionally, our findings can be applied to other transmission lines, enhancing their design and stability in the presence of nonlinear wave phenomena.
\section*{Acknowledgments}
NS wishes to thank DST-SERB, Government of India for providing National Post-Doctoral Fellowship under Grant No. PDF/2023/000619. K.M. acknowledges support from the DST-FIST Programme (Grant No. SR/FST/PSI-200/2015(C)). The work of MS forms a part of a research project sponsored by Council of Scientific and Industrial Research (CSIR) under the Grant No. 03/1482/2023/EMR-II.

\end{document}